\documentclass[12pt]{article}

\pdfoutput=1

\usepackage{amsmath,amssymb,amsfonts,amscd,mathrsfs}
\usepackage{xcolor}
\definecolor{darkblue}{rgb}{0.1,0.1,.7}
\hypersetup{  colorlinks, linkcolor=darkblue, citecolor=darkblue, urlcolor=darkblue, linktocpage}
\usepackage[]{graphicx}
\usepackage{geometry}
\usepackage[margin=10pt,font=small,labelfont=bf]{caption}
\usepackage{ifthen}
\usepackage{tikz}
\usepackage{subcaption}
\usepackage{booktabs,multirow}
\usepackage{hhline}
\usepackage{tablefootnote}

\usepackage{ragged2e}
\usepackage{array}
\newcolumntype{L}[1]{>{\raggedright\let\newline\\\arraybackslash\hspace{0pt}}m{#1}}
\newcolumntype{C}[1]{>{\centering\let\newline\\\arraybackslash\hspace{0pt}}m{#1}}
\newcolumntype{R}[1]{>{\raggedleft\let\newline\\\arraybackslash\hspace{0pt}}m{#1}}

\usepackage{dsfont} 

\usepackage{booktabs}

\usepackage{titlesec}
\titleformat*{\section}{\large\bfseries}
\titleformat*{\subsection}{\normalsize\bfseries}
\titleformat*{\subsubsection}{\normalsize\it}
\titleformat*{\paragraph}{\normalsize\bfseries}
\titleformat*{\subparagraph}{\normalsize\bfseries}

\def\eps{\epsilon}

\newcommand{\beq}{\begin{equation}} 
\newcommand{\eeq}{\end{equation}}

\def\ge{\geqslant}

\def\geq{\geqslant}
\def\leq{\leqslant}
\newcommand{\diffop}[2]{\ifthenelse{\equal{#2}{1}}{\frac{\mrm{d}}{\mrm{d} #1}}{\frac{\mrm{d}^#2}{\mrm{d} #1^#2}}}

\newcommand{\mrm}[1]{{\mathrm #1}}

\usepackage[normalem]{ulem}

\newcommand{\be}{\begin{equation}}
\newcommand{\ee}{\end{equation}}
\newcommand{\bea}{\begin{eqnarray}}
\newcommand{\eea}{\end{eqnarray}}
\newcommand{\eq}[1]{Eq.~(\ref{#1})}
 \def\om{\omega}
  \def\th{\theta}

\let\Im\undefined
\DeclareMathOperator{\Im}{Im}

\makeatletter
\newcommand*{\letterdef@}{}
\newcommand*{\letterdef}[3]{%
  \def\letterdef@##1{\expandafter\newcommand\csname #1\endcsname{#2{##1}}}%
  \@tfor\@tempa :=#3\do{\expandafter\letterdef@\expandafter{\@tempa}}}
\makeatother
\letterdef{bb#1}{\mathbb} {CHPRZ}    
\letterdef{c#1}{\mathcal}{ABCDEFGHIJKLMNOPQRSTUVWXYZ} 
\letterdef{rm#1}{\mathrm} {dDF} 
\letterdef{bf#1}{\mathbf}{pqkxyz}

\usepackage{accents}
\newlength{\dhatheight}

\numberwithin{equation}{section}
\usepackage[tocgraduated]{tocstyle}

\begin{document}

\vspace*{-.6in} \thispagestyle{empty}
\begin{flushright}
\end{flushright}
\vspace{1cm} {\Large
\begin{center}
\textbf{$\mathbf{S}$-matrix bootstrap for resonances}
\end{center}}
\vspace{1cm}
\begin{center}

{\bf  N.~Doroud and J.~Elias Mir\'o}\\[2cm] 
{
 SISSA/ISAS and INFN, I-34136 Trieste, Italy\\
}
 
\vspace{1cm}
\end{center}

\vspace{4mm}

\begin{abstract}

We study the $2\rightarrow 2$ $S$-matrix element of a generic, gapped and Lorentz invariant QFT in $d=1+1$ space time dimensions. We derive an analytical  bound on the coupling of the asymptotic states to unstable particles (a.k.a. resonances) and  its physical implications. This is achieved by exploiting the connection between  the $S$-matrix phase-shift and the roots of the $S$-matrix in the physical sheet. 
We also develop a  numerical framework to recover the analytical bound as a solution to a numerical optimization problem. 
This later approach can be generalized to $d=3+1$ spacetime dimensions.

 \end{abstract}
\vspace{.2in}
\vspace{.3in}
\hspace{0.7cm} \vfill \begin{flushright} April 2018 \end{flushright}

\newpage

\setcounter{tocdepth}{2}

{
\tableofcontents
}

\section{Introduction}

Deriving  the phenomenological  implications of strongly  coupled Quantum Field Theories (QFT) is hard.
Any new idea or approach to inspect such strongly coupled regime deserves to be scrutinized. 
The recent progress on the numerical conformal bootstrap \cite{Rattazzi:2008pe,Rychkov:2009ij,Caracciolo:2009bx,ElShowk:2012ht,Kos:2014bka} -- reviving the successful $d=1+1$ conformal bootstrap \cite{Ferrara:1973yt,Polyakov:1974gs} -- 
 has lead to a revision of the closely related $S$-matrix bootstrap 
 \cite{Paulos:2016fap,Mazac:2018mdx}.

The old analytical $S$-matrix bootstrap approach lost momentum with the advent of QCD and due to the difficulties of dealing with the analytic properties of the $S$-matrix in $d=3+1$ spacetime dimensions.\footnote{See ref.~\cite{wein} for a testimony.} For a compendium of results on the analytic properties of the $S$-matrix see for instance  \cite{Eden:1966dnq}. In the present context \emph{bootstrap} is synonymous to an axiomatic approach, where out of few physical assumptions one extracts general consequences for physical observables. For the $S$-matrix bootstrap approach, the input assumptions are those of quantum mechanics, special relativity and assumptions on the spectrum of particles encoded through analytic properties of the $S$-matrix elements. 

Lately, there has been a number of interesting results within the $S$-matrix bootstrap approach \cite{Paulos:2016fap,Paulos:2016but,Paulos:2017fhb}. 
The key aspects that paved the way for these developments have been to, firstly, identify an interesting and simple enough question that the bootstrap approach can answer and, secondly, the development of a numerical approach to answer the question in general spacetime dimensions. 
Specifically, ref.~\cite{Paulos:2016but} found a rigorous analytical upper bound on the coupling between asymptotic states of the $S$-matrix in $d=1+1$ dimensions.  The existence of such upper bound was expected in higher dimensions and was demonstrated in $d=3+1$ by means of a numerical approach ref.~\cite{Paulos:2017fhb}. 

Exploring the space of consistent $S$-matrices in $d=3+1$ has lots of potential applications for particle physics. For a realistic set up though we would like to study $S$-matrices that feature unstable resonances. The main purpose of this work is to take the first steps towards  developing this theory. 
The present work is entirely in $d=1+1$ and we focus on the $2\rightarrow 2$ $S$-matrix element of the single stable particle of the theory. These simplifying assumptions will allow us to derive a number of analytical results and intuition that is important before attacking the analogous  problem in $d=3+1$. 

Section \ref{pwp} is mostly review and discussion of  the analytic properties of the $S$-matrix.  Section \ref{main1} contains the main result, a bound on the $2\rightarrow 2$ $S$-matrix elements that feature unstable resonances. We discuss the interpretation of this bound and the implications for the spectrum of resonances. 
In section \ref{nums}  we perform a numerical study that matches  the analytical derivations  of section \ref{main1}. Crucially, the numerical approach presented in section \ref{nums} admits a   generalization to $d=3+1$ dimensions. Finally, we conclude and outline possible directions to develop.  

\section{Analytic properties of the $S$-matrix}
 \label{pwp}
 
The main focus of this  paper is to study the space of consistent $S$-matrices in $d=1+1$ spacetime dimensions. For simplicity we restrict our attention to theories with only a single stable particle of mass $m$. This is not crucial and the assumption can be relaxed on later studies. We focus on the elastic $2\rightarrow 2$  $S$-matrix element
\begin{equation}
\label{smat1}
	\langle  p_1, p_2 |\hat S| p_3, p_4 \rangle \equiv\mathds{1} \, S(p_i) \, , 
\end{equation}
where $\mathds{1}=\langle  p_1, p_2  | p_3, p_4 \rangle$ captures the kinematical information. All the interesting physics is encoded in the Lorentz scalar $S$ which is a function of the Mandelstam variable $s=(p_1+p_2)^2$.  Note that in two spacetime dimensions the scattering is along a line and thus there is no scattering angle. Consequently either  of the Mandelstam variables $t=(p_1+p_3)^2$ or $u=(p_1+p_4)^2$ must vanish, which together with the kinematical constraint $s+t+u=4m^2$, imply that the function $S$ in \eq{smat1} is only a function of a single variable $s$. This  function  is further constrained by crossing symmetry
\be
S(s)= S(4m^2-s) \, ,  \label{cross1}   
\ee
\emph{i.e.} it is symmetric under the exchange of the $s$ and $t$ channels (or equivalently between the $s$ and $u$ channels).
In the rest of this section we will review the analytic properties of $S(s)$.

\begin{figure}[t]
\be
\begin{minipage}[h]{0.1\linewidth}
\includegraphics[scale=.475]{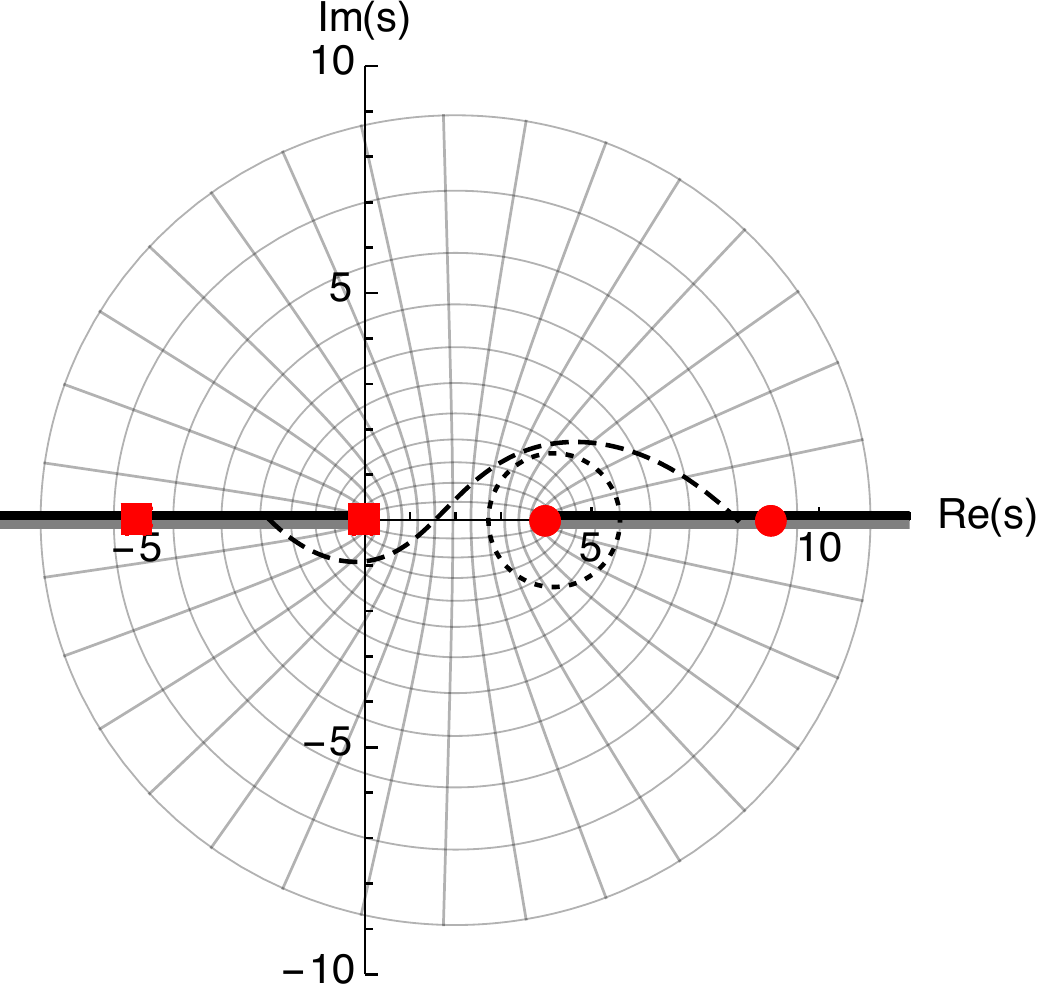}
   \end{minipage} 
   \hspace{3.7cm}
   \begin{minipage}[h]{0.1\linewidth}
\includegraphics[scale=.4]{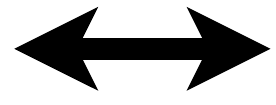}
   \end{minipage}
    \hspace{-.1cm}
   \begin{minipage}[h]{0.1\linewidth}
  \includegraphics[scale=.475]{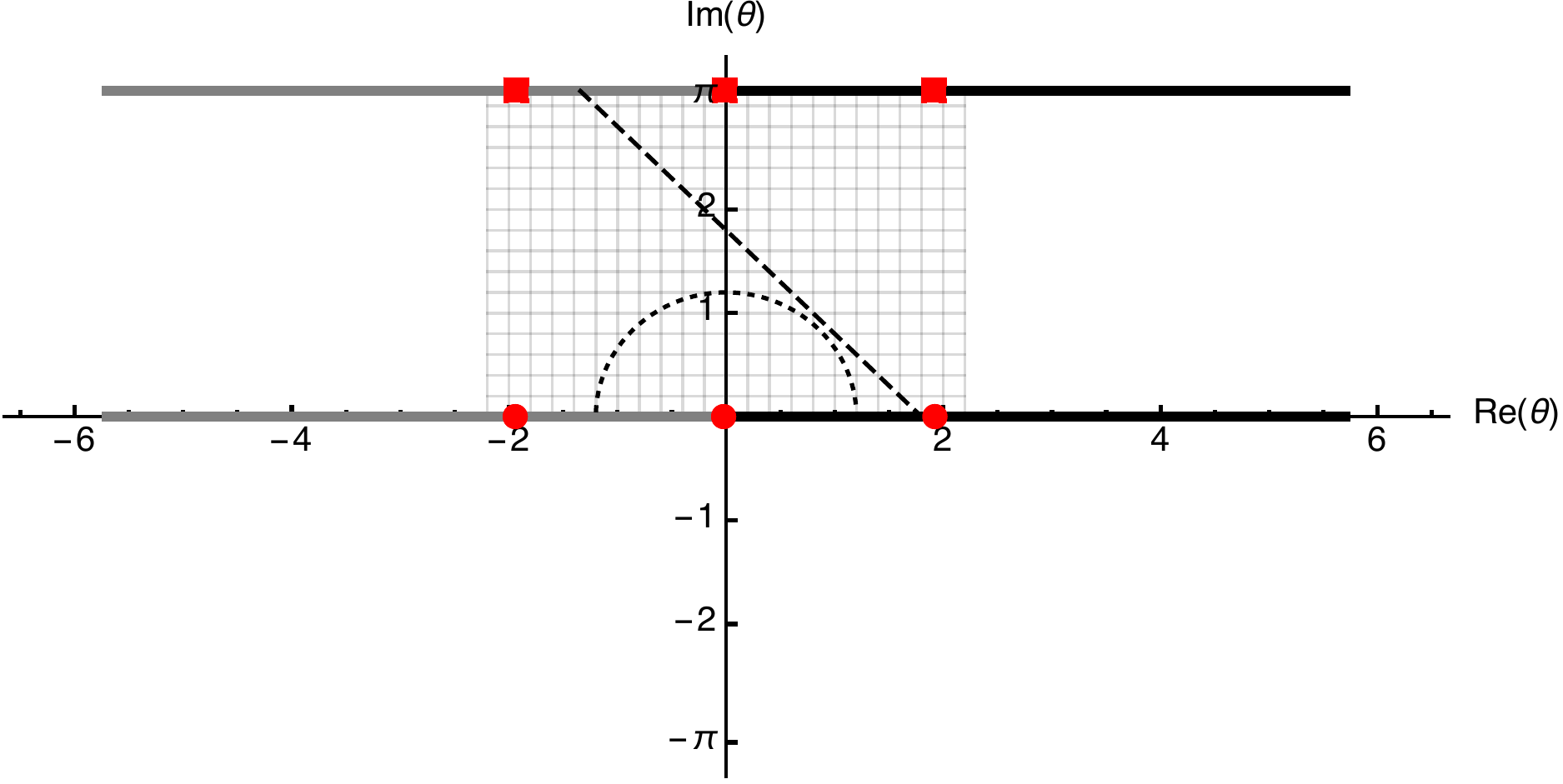} 
   \end{minipage} \hspace{3cm} \nonumber
       \hspace{4cm}
\ee
\caption{ Illustration of the conformal map in \eq{conf1}. The complex $s$-plane, the left plot, is mapped into the complex  $\th$-strip $\Im \th \in (0,\pi)$, right plot. We have also depicted the mapping of  a dashed curve, a dotted curve and a gray grid. \label{fcon} } 
\end{figure}

Consider the analytical continuation of $S(s)$ into the complex $s$-plane. Generically the function $S(s)$ has branch point singularities at the  minimal values of $s$ where  the process $2\rightarrow n$ is kinematically allowed. For positive $s$, the lowest such branch point is at $s=(2m)^2$, the two-particle branch point. Crossing symmetry (\ref{cross1}) implies the presence of a corresponding branch point at $s=0$.
Generically, in the absence of extra symmetries forbidding particle production,  infinitely many branch points are expected on the real line at the minimal values  where higher-particle production is kinematically allowed. 
We have illustrated the branch points at $s=4m^2,9m^2$ (red circles) and the crossing related $s=0,-5m^2$ (red squares) in  the left plot of Fig.~\ref{fcon}.
The physical $S$-matrix is obtained in the limit 
\be
\lim_{\eps\rightarrow 0^{+}}S(s+i\eps) \, , \label{defS}
\ee
namely by approaching the real line from above without encircling any such branch points. 
The physical $s$-plane is defined as the trivial analytical continuation of \eq{defS} without encircling any branch point. Pictorially, it consists of the full complex plane minus the cuts on the real axis, see Fig.~\ref{fcon}. 
The two key analytical assumptions on $S(s)$ are that all the singularities of the physical $s$-plane consist only of branch points on the real line;~\footnote{We assume a $Z_2$ symmetry forbidding a cubic self-interaction of the stable particle. Such cubic self-interaction  would lead to poles at  $s=m^2, 3m^2$.�\label{ft1} }
and that along the real axis and below the two-particle threshold $S(s)$ is a real function. Thus the analytic continuation of $S(s)$ satisfies
\be
S^*(s)=S(s^*)  \, , \label{analy} 
\ee
which is often referred to as \emph{real analyticity}.~\footnote{Let us note that in \eq{analy} we have assumed that the $S$-matrix theory is invariant under space parity. The general condition for the two-body $S$-matrix is Hermitian analyticity  $S^*_{ij}(s)= S_{ji}(s^*)$ which reduces to real analyticity only for parity invariant theories~\cite{Miramontes:1999gd}.}

The last   property of $S(s)$ follows from unitarity of the full $S$-matrix, implying  the following constraint on the $2\rightarrow 2$  $S$-matrix element
\be
	S(s +i\eps) S(s-i\eps) = f(s)   \quad \quad \text{with} \quad 0\leq f(s)\leq 1 \label{unit2} \, ,
\ee
and $s>4m^2$. Note that we have used real analyticity to write the modulus as $|S(s+i\eps)|^{2}= S(s +i\eps) S(s-i\eps) $.
Recall that below the inelastic threshold $s_*$ (above which $2\rightarrow n$, with $n>2$, processes are kinematically allowed) and above the two-particle production threshold, unitarity is saturated
\be
f(s)=1 \quad \text{for}\quad \quad 4m^2< s \leq s_* \, .
\ee
Typically the inelastic threshold is at the three-particle production threshold $s_*=9m^2$ or four-particle production threshold $s_*=16m^2$.

Unless explicitly stated otherwise, we will refer to  $S(s)$ as the \emph{$S$-matrix} (instead of \emph{the $2\rightarrow 2$ $S$-matrix element function}). To summarize, the $S$-matrix is assumed to satisfy crossing (\ref{cross1}), 
real-analyticity (\ref{analy}), unitarity (\ref{unit2}) and there are no  singularities in the physical $s$-plane but only branch points on the real line associated with the $2\rightarrow n$ ($n\geq 2$) scattering processes.
In order to further  elucidate the analytical properties of the $S$-matrix we will next  review a particularly simple  $S$-matrix. This will also serve as an excuse to introduce the rapidity variable $\th$ which we will use in the rest of the paper.

 \subsection{The $\th$-strip}
 \label{introth}

 Consider the following classic QFT example in $d=1+1$: the Sine-Gordon $S$-matrix element for the scattering of the lightest breather is given by~\cite{Zamolodchikov:1976uc,Arefeva:1974bk}
\be
S_{SG}(s)= \frac{\sqrt{s}\sqrt{4m^2-s}+m_1\sqrt{4m^2-m_1^2}}{\sqrt{s}\sqrt{4m^2-s}-m_1\sqrt{4m^2-m_1^2}} \, , \label{sg1}
\ee
where $s$ is the Mandelstam variable. The function $S_{SG}(s)$ has a pole at $s=m_{1}^{2}$, the mass of the next-to-lightest breather. The matrix element $S_{SG}(s)$ has branch points at $s=0,\, 4m^2$ associated with the two-particle production threshold. These are square-root branch points and can be  resolved by the conformal map
\be
s(\th)=4m^2 \cosh^2(\th/2) \, . \label{conf1}
\ee
\eq{conf1} maps the entire physical  $s$-plane minus the cuts on the real line  into the strip 
 \be
 \Im\th\in(0,\pi) \, , \label{phstrip}
 \ee
  which we will refer to as the \emph{physical strip}. The transformation is illustrated in Fig.~\ref{fcon}. The two-to-two $S$-matrix element (\ref{sg1}) in the $\th$-strip is given by
\be
S_{SG}(\th) \equiv S_{\alpha}=  \frac{\sinh \th-\sinh\alpha}{\sinh\th + \sinh\alpha} \, , \label{sg2}
\ee
where $\sinh\alpha=-i\,  m_1/m \sqrt{1-m_1^2/(4m^2)}$, and we defined $S_\alpha(\th)$ for later use.~\footnote{The function $S_\alpha(\th)$ is commonly called a Coleman-Dalitz-Dyson (CDD) factor.}
\eq{sg2}  is analytic at the points $\th=0$ and $\theta=i\pi$, corresponding to the original branch points of \eq{sg1} at $s=4m^2$ and $s=0$ respectively. 
The second Riemann sheet  of (\ref{sg1}) reached by traversing a branch cut stemming from the two-particle branch points at $s=0,4m^2$ is mapped into $\Im\th\in(-\pi,0)$. Note also that    the lines $\th=-i\pi$ and $\th=i\pi$ are identified; $\Im\th\in[-\pi,\pi)$ is the fundamental domain of \eq{sg2}, which is periodic under $\th\sim\th+2\pi i $.

The Sine-Gordon theory is very special as it is an integrable QFT. It follows that there is no particle production and the full $S$-matrix factorizes into the product of $2\rightarrow 2$ matrix elements. Consequently $S_{SG}(\th)$ is a meromorphic -- and thus single valued -- function in the $\th$-strip $\Im\th\in[-\pi,\pi)$.  
For a generic non-integrable QFT however, one has branch points at the inelastic thresholds $s=\{(3m)^2,\, (4m)^2,...\}$ where the matrix elements $S_{2\rightarrow 3, 4 \dots}$ are switched on.
Those are mapped into the real line of the $\th$-strip. As depicted in Fig.~\ref{fcon}, they appear both in the positive and negative real axis of the $\th$-strip because they can be reached from both Riemann sheets associated to the two-particle branch point. 
 
As we have seen, the branch point at the two-particle production threshold in the particular example \eq{sg1} is two-sheeted. It turns out that this feature is more general and extends to non-integrable $S$-matrices, see appendix \ref{branchpoint} for further details. The results of this paper however do not make use of the nature of any of the branch points in the physical $s$-plane.
 
 In $d=1+1$ dimensions  $\th$ has the physical interpretation of being the rapidity difference of the incoming particles $\th\equiv \th_2-\th_1$ where $p_{i}=(m \cosh\th_i, \, m\sinh\th_i)$. The $S$-matrix literature in $d=1+1$ dimensions commonly uses this variable. Thus, in the rest of the paper we will consider the $S$-matrix as a function $S(\th)$ (this is however   not crucial and all the results below can be reformulated in the $s$-plane). For completeness, let us recall that crossing symmetry (\ref{cross1})  in the $\th$-strip implies
 \be
 S(\th)= S(i\pi-\th) \, ,
 \ee
 real analyticity  (\ref{analy}) reads
 \be
 S^*(\th)=S(-\th^*) \, , 
 \ee
and  unitarity (\ref{unit2}) in the $\th$-strip reads
 \be
 S(\th)S(-\th)= f(\th)       \, ,\label{unit3}
 \ee
where $ 0\leq f(\th)\leq 1$ for real $\th$.

 \section{Unstable resonances}\label{main1}

Our goal is to study QFTs with unstable particles or resonances. In this section, we first present the operational definition of a resonance before deriving a bound for the $S$-matrices that feature resonances. Finally, we discuss the interpretation of the bound in the context of a weakly coupled QFT. 
 
\subsection{What are unstable resonances?}
\label{class}

In perturbation theory unstable particles are often associated with complex poles. These poles lie on higher Riemann sheets that can be reached by traversing the multi-particle branch cuts along the real line in the $\theta$-plane.\footnote{See  appendix \ref{zp} for  an explicit example.}
The distinguishing feature of such singularities is that they  lead to pronounced variations of  the phase of the $S$-matrix evaluated along the real line.
Thus, we define a resonance as an abrupt change in the phase of the $S$-matrix 
\be
 \text{Re}\,  2\delta(\th) \quad \text{where} \quad 2i\delta(\th)\equiv\log S(\th) \,  , \label{psx}
\ee
without any reference to poles in higher Riemann sheets.

Abrupt variations of the phase of the $S$-matrix typically signal the presence of poles or zeros of the $S$-matrix in the complex plane and it is up to us to classify such pronounced features of the $S$-matrix. 
Of particular interest is when the  phase of $S(\th)$ abruptly increases by $2\pi$ continuously and monotonically in $\th$. Such $2\pi$ phase-shifts 
 stem from the presence of a pair of zeros in the $S$-matrix $S(\th)$ in the physical strip $\Im\th\in(0,\pi)$. In fact, each zero   in the physical $\th$ strip contributes with an $i\pi  $ to the total $S$-matrix phase shift~\footnote{See appendix  \ref{deri} for the derivation of \eq{totalps} -- this is the relativistic analog of Levinson's theorem, see for instance chapter XVII of \cite{Landau:1991wop}.}
\be
2 \int_{-\infty}^\infty d\th \,  \partial_\th  \delta(\th) =\sum_\text{zeros} \pi   \, . \label{totalps}
\ee
Due to crossing symmetry, the zeros $\th_i$ of $S(\th)$ come in pairs related by crossing $S(\th_i)=S(i\pi-\th_i)=0$. 
In addition, by real analyticity,   the roots are also pairwise related by complex conjugation $S^*(\th_i)=S(-\th_i^*)=0$. 
In many physically relevant $S$-matrices one finds  an approximate $2\pi$ change of the phase   in a bounded span  $\th\in[\th_{\circ}-\gamma,\th_{\circ}+\gamma]$
\be
2\Delta \delta \equiv2\int_{\th_{\circ}-\gamma}^{\th_{\circ}+\gamma} \partial_\th \delta(\th)\,  d\th \approx 2\pi \, ,  \label{ps2pi}
\ee
where $\th_\circ = \text{Re}\, \th_i$ and $\gamma \sim2 \Im \th_i$, the exact choice of the resonance region $\th_\circ\pm\gamma$ is somewhat arbitrary. This is the kind of resonances that we are interested in this paper.

\begin{figure}[t]\centering
\includegraphics[scale=.575]{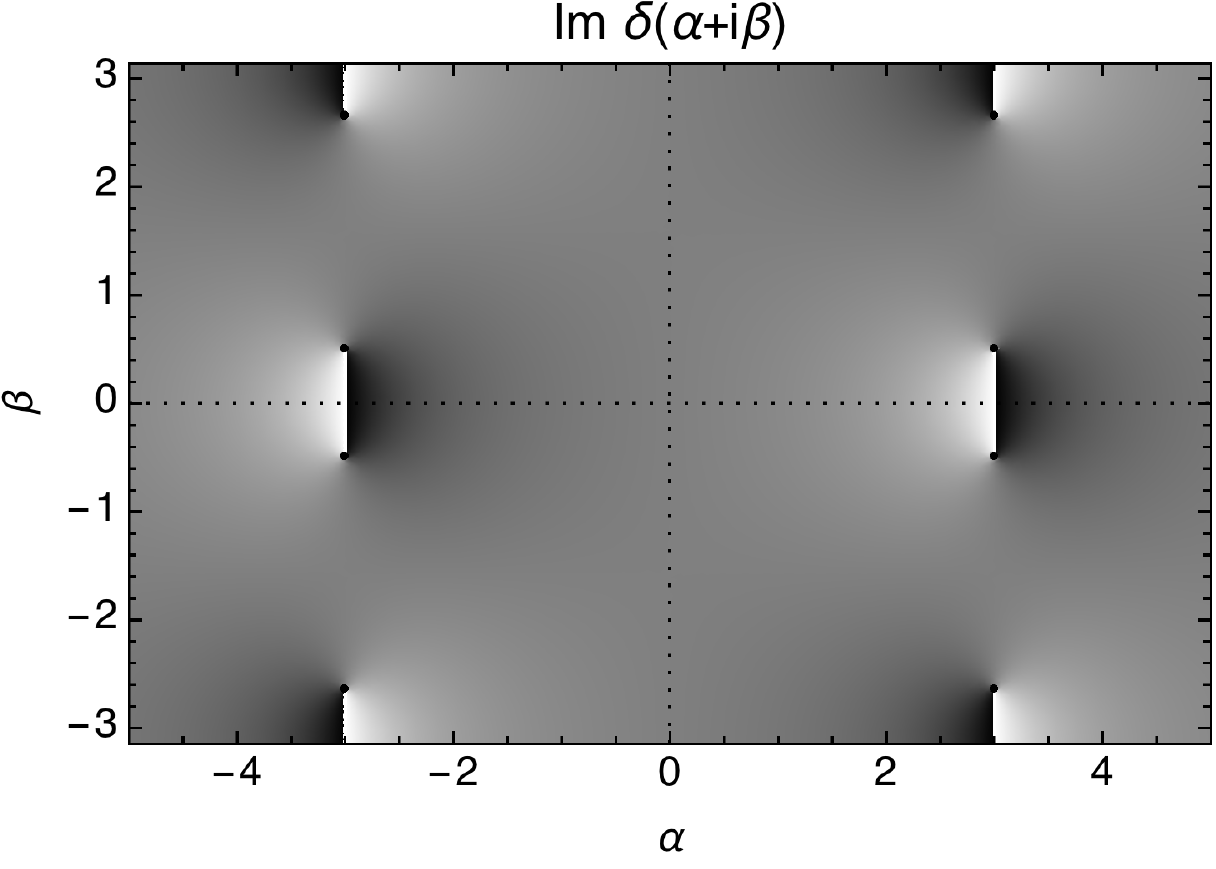}\hspace{1cm}\includegraphics[scale=.569]{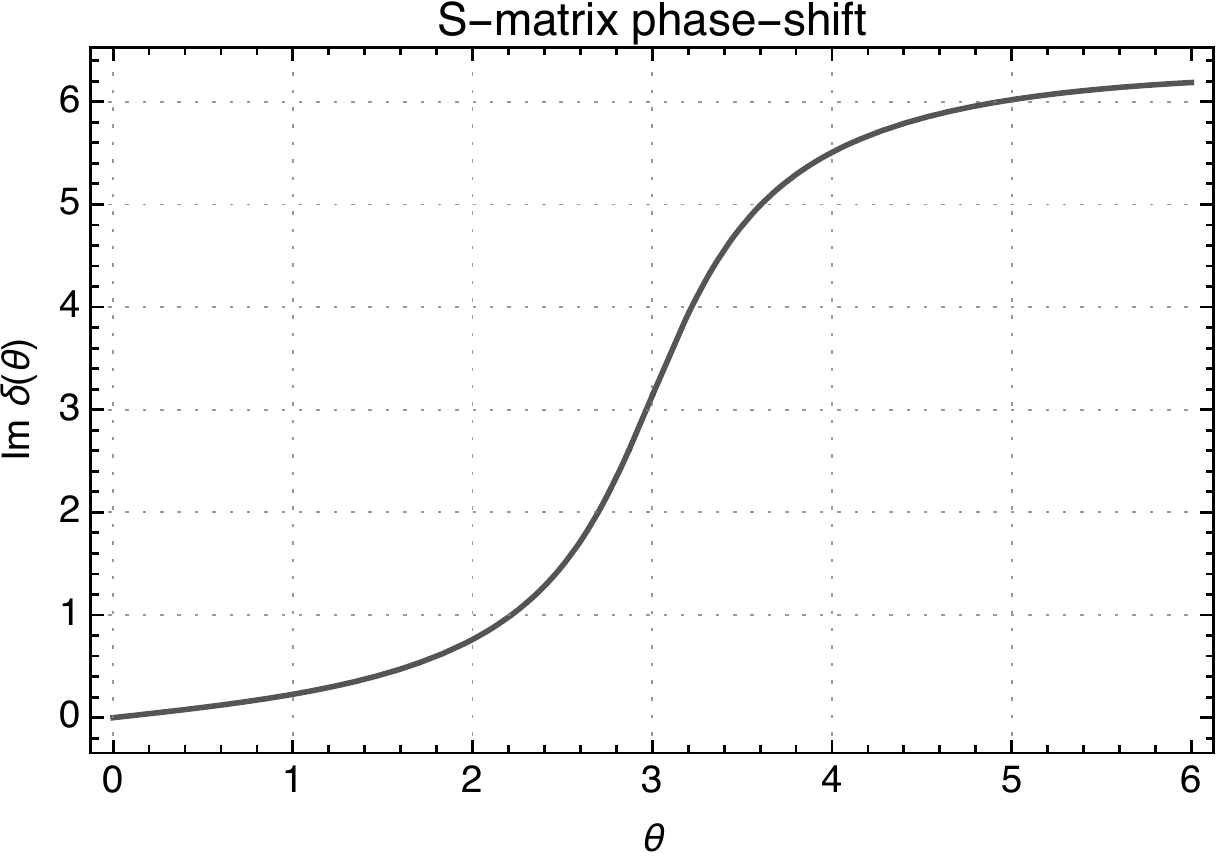}
\caption{\textbf{Left:} Section of the complex plane of the phase of \eq{scatex}, the lines $\delta(i\pi)\sim\delta(-i\pi)$ should be indentified. \textbf{Right:} phase-shift of \eq{scatex} localized around the position of the branch points generated by the zeros and poles of $S_\text{ex}(\th)$.  \label{figex}} 
\end{figure}

\subsubsection{Further comments}

For a long-lived unstable particle associated to a long time delay, the $2\pi$ phase-shift is highly localized and $\Im\th_i \ll 1$. The zeros in the physical strip are accompanied by poles which are hidden behind the multi-particle branch cuts. In terms of the $\th$ variable the $S$-matrix  
 behaves as
\be
e^{2i\delta(\th)}\sim e^{2i \delta_0} \, \frac{\th-\th_i}{\th-\th_i^*} \, , 
\ee
for $\th$ close to  $|\th_i|$.
 The zeros and poles of $S(\th)$ result in branch points of     $\delta (\th)$. Pictorially, this leads to a  branch cut that ``cuts" the real line. Then, when evaluating $2\delta(\th)$ along the real line   the $2\pi$ phase-shift is a consequence of changing Riemann sheet of the logarithm. 

As an illustration, consider the following function 
\be
S_\text{ex}(\th)=S_{\alpha}(\th) S_{-\alpha^*} (\th)  \, ,\label{scatex}
\ee
where $ \Im \alpha>0 $ and $S_\alpha(\th)$ was defined in \eq{sg2}. The function $S_\text{ex}(\th)$ is a realistic  $S$-matrix  element because it  satisfies the unitary equation, crossing symmetry and it is real analytic.\footnote{Note however that a generic product of CDD factors leads to a finite volume  spectrum $E_i(R)$ with branch point singularities at finite volume \cite{Mussardo:1999aj,Smirnov:2016lqw}.}
 $S_\text{ex}(\th)$ has zeros at $\th=\alpha,-\alpha^*$ and poles at complex conjugate points  as required by unitarity $S_{ex}(\th)S_{ex}(-\th)=1$. 
The left plot in Fig.~\ref{figex} shows a section of the fundamental domain of the complex plane of $\Im \delta(\th)$ that includes the zeros and poles of $S(\th)$. We have depicted branch cuts connecting the zeros on the physical strip with the poles on the lower stip $\th\in(-\pi,0)$. The branch cuts intersect the real line along which the physical $S$-matrix is evaluated. On the right hand side we have ploted $\text{Re } 2\delta(\th)$ on a segment along the real line. As $2\delta(\th)$ goes through the region $\th\approx \text{Re}\alpha$, \emph{i.e.} near the location of the branch points, the function  $2i\delta=\log S$ is evaluated in a higher Riemann sheet and the imaginary part is shifted by $2\pi$. In section~\ref{inter} we discuss a perturbative QFT   with the same qualitative picture as the $S$-matrix in \eq{scatex}.

To close up this section, let us insist that in general we will not refer to complex poles of the $S$-matrix. Instead,   we focus on the zeros of $S(\th)$ in the physical $\th$-strip (or  physical $s$-plane), which are in a one-to-one correspondence with each $\pi$ contribution to the total  phase-shift (\ref{totalps}). 
This picture avoids the need to discuss the nature of the branch points of $S(\th)$ on the real line and the analytical continuation of the function $S(\th)$ around such branch points which requires a case by case analysis.  Instead, it only requires the trivial analytical continuation of $S(\th)$ into the physical sheet which, by definition, is always available.
Note also that the operational definition of unstable resonance that we are employing is physically meaningful because the phase-shift is experimentally accessible (it can also be extracted from lattice Monte Carlo simulations \cite{Luscher:1990ck}).

\subsection{A bound on the $S$-matrix of unstable resonances}
 
The two-dimensional $S$-matrix can be written as follows
\be
S(\th)= \prod_j S_{\alpha_j}(\th) \text{ exp}\left(-\int_{-\infty}^{+\infty} \frac{d\th^\prime}{2\pi i }\frac{\log f(\th^\prime)}{\sinh (\th-\th^\prime+i \eps)}\right) \quad \text{for}\quad \Im(\th) \in [0,\pi)    \label{lg}
\ee
where $\eps$ is an arbitrarily small positive parameter and $S_{\alpha_j}(\th)$ denotes a CDD factor defined in (\ref{sg2}):
\be
S_{\alpha}(\th) = \frac{\sinh\th-\sinh\alpha}{\sinh\th+\sinh\alpha}\, . \nonumber
\ee
The set $\{\alpha_j\}$ parametrizes  the position of the zeros and poles  of $S(\th)$ in the physical strip. As written in \eq{lg} the set $\{\alpha_j\}$ may contain repeated elements in order to account for the correct order of the poles and zeros of $S(\th)$.~\footnote{In fact, for our particular physical set up with a single stable particle $\prod_j S_{\alpha_j}(\th)$ has no poles in the physical strip but only zeros.}  The function $S_{\alpha}(\th)$ saturates unitarity $S_\alpha(\th)S_\alpha(-\th)=1$ along the entire real line. The function $f(\th)$ parametrizes the amount of inelasticity, see (\ref{unit3}).  
 \eq{lg} will play a crucial role in our discussion below so let us review its derivation. 

\subsubsection{Discussion of \eq{lg}}

Let us define $\phi(\th)\equiv  2i  \partial_\th \delta(\th)$ and  consider the following dispersion relation
 \be
 \phi(\th) = \oint_{\partial {\cal C}_\th}\frac{d\th^\prime}{2\pi i} \frac{\phi(\th^\prime)}{\sinh(\th^\prime-\th)} \,  , \label{toblow}
 \ee
 where  $\partial {\cal C}_\th$ is a closed contour encircling a region ${\cal C}_\th$ where $\phi(\th)$ is regular, \emph{i.e.}  such that $S(\th)$ is holomorphic and does not vanish in ${\cal C}_\th$.
 Next we apply Cauchy's theorem and  blow the contour  in \eq{toblow} to the boundary of the physical strip.  In doing so, we must subtract the zeros of $S(\th)$ in the physical strip 
 \bea
 \phi(\th) &=&\sum_j\left(\frac{1}{\sinh(\alpha_j-\th)}+\frac{1}{\sinh(i\pi-\alpha_j-\th)}\right) \nonumber\\[.2cm]
 & +&\int_{-\infty}^\infty \frac{d\th^\prime}{2\pi i} \ \frac{\phi(\th^\prime)}{\sinh(\th^\prime-\th)} +\int_{\infty+i\pi}^{-\infty+i\pi} \frac{d\th^\prime}{2\pi i} \ \frac{\phi(\th^\prime)}{\sinh(\th^\prime-\th)}   \,  , \label{blow1}
 \eea
 where  the $\th=0,\, i\pi$ lines are approached from above and below, respectively.  In \eq{blow1}, zeros  come in pairs related by crossing $S(\th)=S(i\pi-\th)$ and  we have dropped the contribution from the contour arcs at infinity. This can be justified by assuming that  $S(\th)$  is polynomially bounded.~\footnote{We need this kind of technical assumption to prove \eq{lg}. However, as discussed in section~\ref{morecom} below,  this assumption is not crucial for the bound on the $S$-matrix that we are about to derive.}
Crossing symmetry implies $\phi(\th+i\pi)=\phi(-\th)$. Therefore, the last integral in \eq{blow1} can be written as $\int_{-\infty}^{\infty} \frac{d\th^\prime}{2\pi i} \ \phi(-\th^\prime)/\sinh(\th^\prime-\th)$ and we are led to
 \bea
 \phi(\th) =\sum_j\left(\frac{1}{\sinh(\alpha_j-\th)}+\frac{1}{\sinh(i\pi-\alpha_j-\th)}\right)-\int_{-\infty}^\infty \frac{d\th^\prime}{2\pi i} \frac{  \partial_{\th^\prime}\log f(\th^\prime) }{ \sinh(\th^\prime-\th)  } \, .  \label{blow2}
 \eea
where we have used $\phi(\th)+\phi(-\th)=\partial_\th\log f(\th)$, by \eq{unit3}. 
Finally,  integrating by parts with respect to $\th^\prime$ the integral in \eq{blow2} and using $S(\th)=\exp \int d\th \phi(\th)$  we are led to \eq{lg}.~

\bigskip

A key point of \eq{lg} is that the roots $\{\alpha_j\}$ of the $S$-matrix   are factored out. 
The factor in \eq{sg2} shows that each zero $\alpha_j$ in the physical strip   has  an accompanying pole  located at  $-\alpha_j$ in the unphysical strip $\Im\th\in (-\pi,0)$. 
Note however, that this observation does not necessarily imply that each zero $\alpha_j$ of the $S$-matrix in \eq{lg} has a pole at $-\alpha_j$.
\eq{lg} only applies in the physical strip. In order to analytically continue \eq{lg} into the unphysical strip  we need to know the nature of the branch point singularities on the real line. If the two-particle threshold branch point at $s=4m^2$ is a square-root singularity (in the Mandelstam $s$-plane), then the conformal map $s=4m^2\cosh^2(\th/2)$ resolves the singularity and we can analytically continue \eq{lg} provided we avoid other possible branch points on the real line. Then we may conclude that $S(\th)$ has poles in the unphysical strip at the same positions as the   factors $S_{\alpha_j}(\th)$. But again, this conclusion is not strictly necessary. 

Even if we can analytically continue the function \eq{lg} across the branch points on the real line, the poles of $S_{\alpha_j}(\th)$ may be canceled by the exponential factor $\sim e^{\int \log f/\sinh}$ in \eq{lg}, which at the same time  can generate poles in unphysical sheets reached by traversing higher particle production  branch cuts not related to the two-particle branch point.

\subsubsection{The bound}
\label{bound}

Consider a generic point  $\th= \tilde \th + i t$, with $\tilde \theta\in \mathbb{R}$ and $t\in(0,\pi)$ in the physical strip.
Then,  the absolute value of the $S$-matrix is given by
\be
|S(\th)|=\prod_j |S_{\alpha_j}(\th)| \text{ exp}\left(\sin t \int_{-\infty}^{+\infty} \frac{d\th^\prime}{2\pi   } \frac{ \cosh(\tilde\th-\th^\prime)  }{|\sinh (\th-\th^\prime)|^2}   \log f(\th^\prime)  \right)  \, , \label{lgo}
\ee
where we have used $ \text{Re}  [i  \sinh^*( \th -\th^\prime) ]=  \sin t \cosh(\tilde\th-\th^\prime) $.
Note that $\log f(\th)\leq0$ in the whole integration domain because $0\leq f(\th^\prime)\leq 1$ on the real line. Therefore we have
\be
|S(\th)|\leq \prod_j |S_{\alpha_j}(\th)|   \ ,  \label{lgp}
\ee 
for   $\th$ in the physical strip.
\eq{lgo} applies in the whole physical $\th$-strip and  in particular it implies 
\begin{equation}
	|S^{\prime}(\alpha_{i})| \leq |S^{\prime}_{\alpha_{i}}(\alpha_{i})| \prod_{j\neq i} |S_{\alpha_{j}}(\alpha_{i})| \, , \label{lgp2}
\end{equation}
at the position of each zero $\alpha_i$ in the physical strip. 

We shall see in section \ref{inter} that there is a direct relation between $|S^{\prime}(\alpha_{i})|$ and the parameter controlling the perturbative expansion, \emph{i.e.} the dimensionless coupling constant. More precisely, we will show that $S^\prime(\alpha_i)$ is proportional to the ratio of the expansion parameter and the square of the width of the resonance. 
Thus, for a fixed width and with an 
abuse of language, we will call
\be
S^\prime(\alpha_i) \text{\emph{: coupling to the resonance $\alpha_i$}} \, ,
\ee
without reference to the actual underlying coupling constants of the possible Lagrangian description.

\subsubsection{Further comments on the bound}
\label{morecom}

The derivation of \eqref{lgp} presented above  explicitly accounts for inelasticity $f(\theta)$ and knowledge of   $f(\theta)$ results in a stronger bound than (\ref{lgp}). Note, however, that the bound can be obtained in a more general setting as follows. Consider the nowhere-vanishing function 
\be
h(\th)=S(\th)/\prod_j S_{\alpha_{j}}(\th) \, , 
\ee 
where the product in the denominator runs over all zeros of $S(\theta)$ with the appropriate order. By construction, $h(\theta)$ is a holomorphic function in the physical strip and is bounded on the boundaries $\Im\theta_{b}=0, \,  \pi$, since $|S_{\alpha_k}(\th_b)|=1$ and $|S(\theta_{b})|\leq1$. Therefore, by the Hadamard three-lines theorem, $|h(\theta)|$ is bounded in the physical strip by its value at the boundary and we are led to  \eq{lgp}.  
In refs.~\cite{Creutz:1973rw,Paulos:2017fhb} a similar argument is used to bound the residue of the poles of $S(\th)$ on the $\th\in[0,i\pi)$ segment in the physical strip which are associated with stable particles.
\bigskip

 The simple derivation of  (\ref{lgp}) given above does not require the $S$-matrix to admit a representation of the form \eqref{lg}. As an example of such an $S$-matrix of broad interest, consider
\be
S_{g }(\th)=e^{2g\sqrt{ \cosh^2(\th/2)}\sqrt{1-\cosh^2(\th/2)}} \equiv e^{ig\sinh \th}\, , \label{gravS}
\ee
where $g\geq 0 $. The $S$-matrix \eqref{gravS} does not admit a representation of the form \eqref{lg} with \emph{finitely many factors} of $S_{\alpha}$. In fact one can check the the above $S$-matrix can be obtained in the limit where we have an infinite product of $S_\alpha$ factors \cite{Smirnov:2016lqw}:
\be
\label{prgravS}
e^{i g \sinh\th}=   \lim_{n\rightarrow\infty}  (-1)^{n}\prod_{j=1}^{n} S_{\alpha}(\theta)
\ee
where $S_{\alpha}$ is given by \eqref{sg2} with $\sinh\alpha = 2in/g$. 
The $S$-matrix $S_{g}(\th)$ has infinitely many phase-shifts of the type in \eq{ps2pi} that can be interpreted as resonant particles~\cite{Dubovsky:2012wk}. This is easily explained from the infinite-product representation above: each $S_\alpha$ factor accounts for a pair of (simple) zeros in the physical strip\footnote{Higher order zeros can be factorized by point-splitting, $S_{\alpha}^{n} = \prod_{j=1}^{n} S_{\alpha+j \epsilon}$, where $\epsilon\sim e^{-n}$.} giving rise to a phase shift of $2\pi$. In the limit $n\rightarrow\infty$ in \eqref{prgravS} we end up with an infinite number of coincident zeros at $\theta=\infty+ i\pi/2$, and the total (integrated) phase shift is infinite.~\footnote{Other examples with infinitely many resonances are the elliptic (doubly periodic) $S$-matrix such as ref.\cite{Zamolodchikov:1979ba,Mussardo:1999ee}. }
The $S$-matrix in (\ref{gravS}) can be viewed as an  integrable deformation \cite{Smirnov:2016lqw,Dubovsky:2012wk,Cavaglia:2016oda} corresponding to the upward flow generated by certain irrelevant operators (in the RG sense). Moreover, a special limit of \eqref{gravS}, namely  $\lim_{g\rightarrow 0 }S_{g}$, appears in the context of the effective string description of Yang-Mills flux tubes \cite{Dubovsky:2014fma}.

\begin{figure}[t]\centering
\includegraphics[scale=.378]{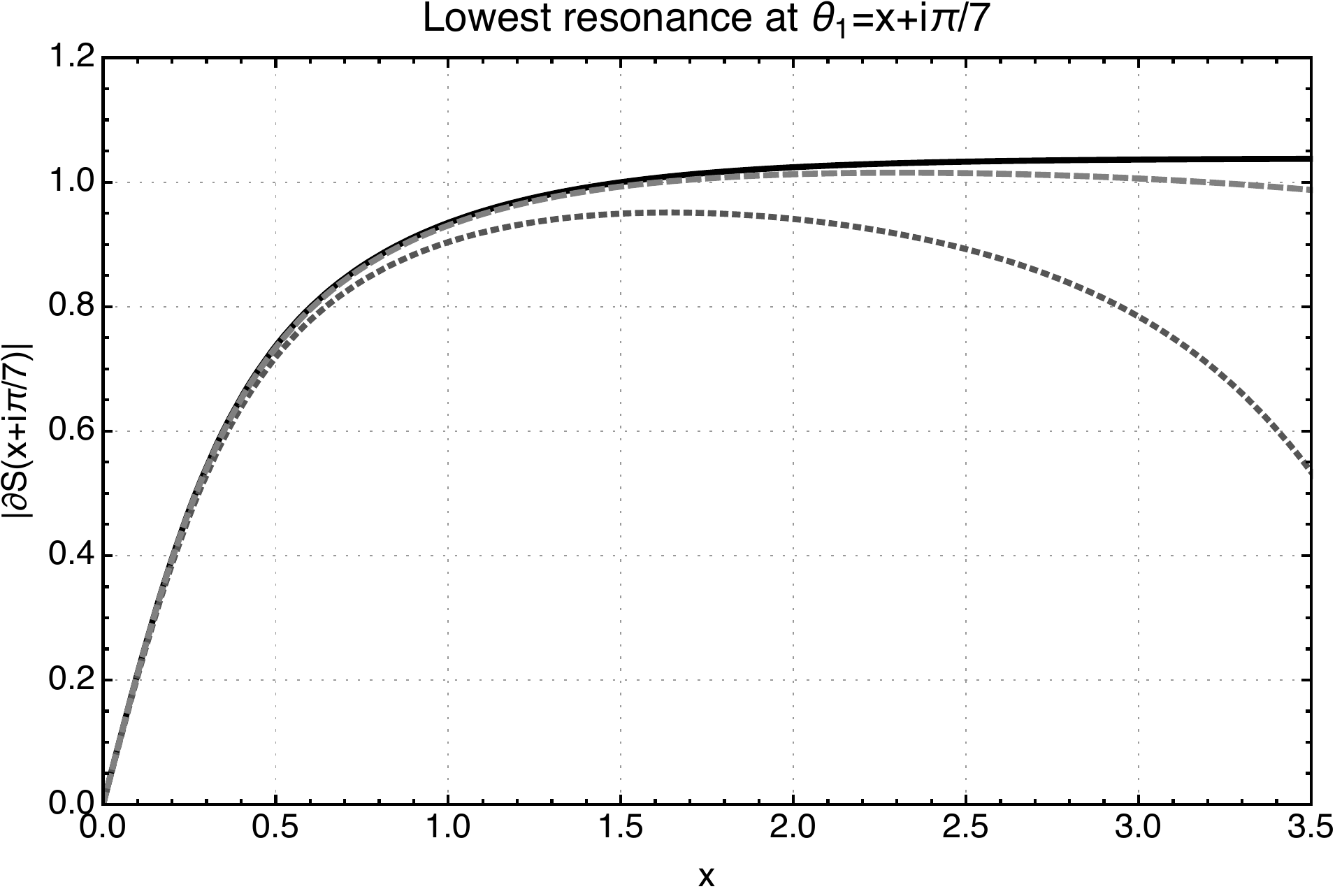} \quad \quad \quad \includegraphics[scale=.3645]{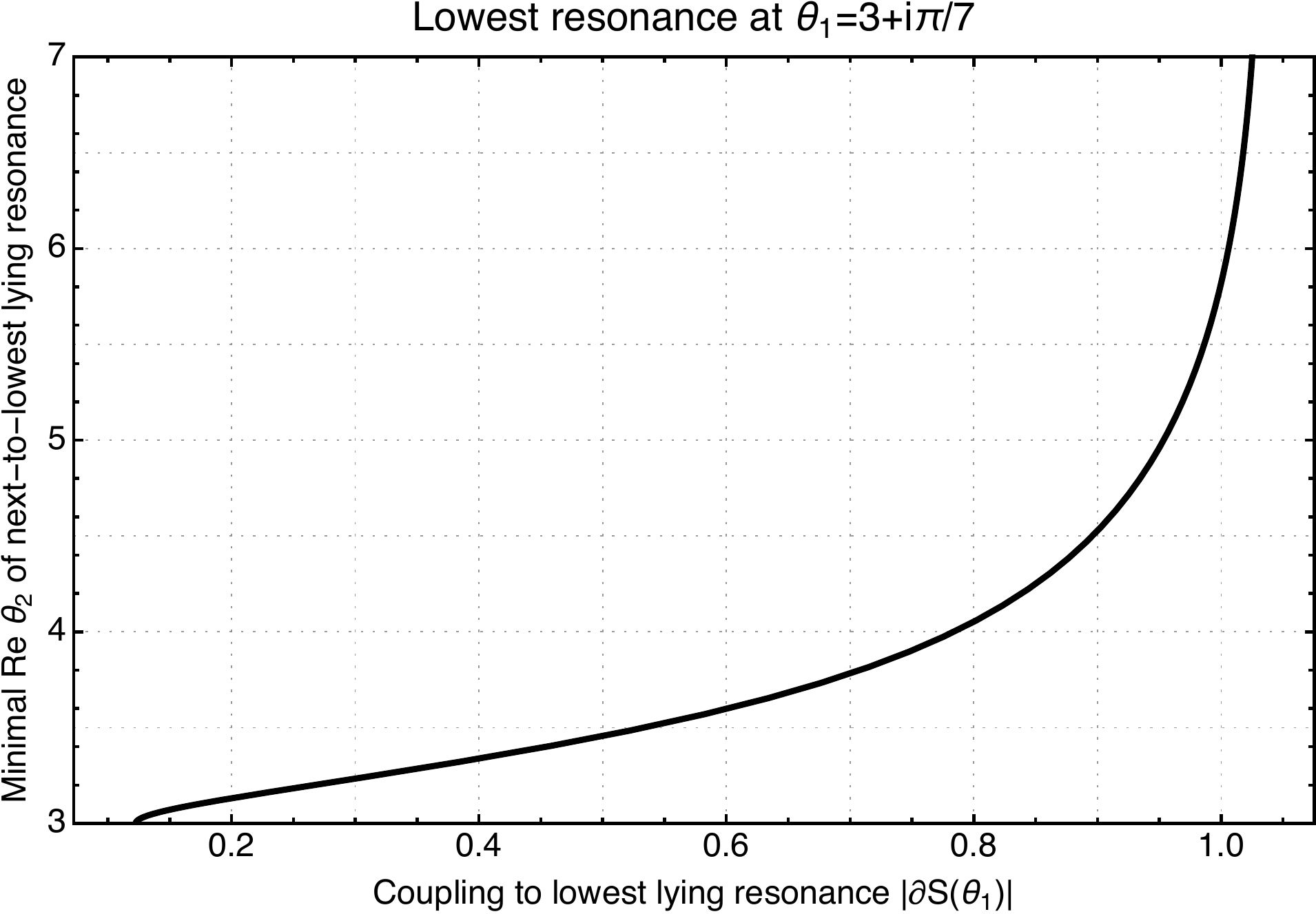}
\caption{In the left plot, maximal coupling to the lowest mass resonance at $\th_1=x+i\pi/7$ for an $S$-matrix with a single resonance (solid black), and an $S$-matrix with a second resonance at $\th_2=6+i\pi/9$ (dashed light gray)  and $\th_2=4+i\pi/9$ (dotted darker gray). On the right, minimal value of $\text{Re}\th_2$ as a function of the coupling to the $\th_1$ resonance for an $S$-matrix with two resonances.  \label{fdec}} 
\end{figure}

\bigskip

 \eq{lgp} implies that an $S$-matrix \emph{with at least $n$ zeros} at $\{\alpha_j\}$ has a magnitude  less than or equal to $\prod_{j=1}^{n}|S_{\alpha_j}(\th)|$,  with  $|S_{\alpha}(\th)|\leq 1$ for  $\th$ in the physical strip. Therefore, we do not  necessarily  need to know the spectrum of unstable resonances up to arbitrarily high energy in order to obtain a meaningful bound. Additional knowledge of UV resonances makes the bound more stringent.
This observation is  key for the bounds in (\ref{lgp}-\ref{lgp2}) to be  sensible from an effective low energy physics standing point where we do not necessarily want to commit to a particularly detailed spectrum of UV resonances beyond a certain energy cutoff. 
To illustrate this point we have plotted $|S^{\prime}(\theta_{1})|$ as a function of $x\equiv\text{Re}\theta_{1}$ for the S-matrix $S_{1} = S_{\theta_{1}}S_{-\theta_{1}^{*}}$ (black solid line) in the left plot in Fig.~\ref{fdec}. Any other theory which features resonances at $\{\theta_{1},\,-\theta_{1}^{*}\}$ has a coupling $|S^{\prime}(\theta_{1})|$ which falls below this line. For comparison, the same plot depicts $|S^{\prime}(\theta_{1})|$ for the S-matrix $S_{2} = S_{\theta_{1}}S_{-\theta_{1}^{*}}S_{\theta_{2}}S_{-\theta_{2}^{*}}$ featuring a second pair of resonances at $\{\theta_{2},\,-\theta_{2}^{*}\}$. The dotted line depicts $|S_{2}^{\prime}(\theta_{1})|$ for $\theta_{2}=4+i\pi/9$ and the dashed line is plotted with $\theta_{2}=6+i\pi/9$. As can be seen from the plots the closer the resonances are to each other the stricter the bound on $|S^{\prime}(\theta_{1})|$ gets. Thus, the effects of possible further heavy resonances $|\theta_{1}-\theta_{j}| \gg 1$ decouples at low energy.

The latter observation suggests another interesting viewpoint on the bound (\ref{lgp2}): the larger the value of $|S^\prime(\th_1)|$, the larger the minimum mass gap with the nearest resonance $\th_2$.
Namely, the mass gap increases monotonically as $|S^\prime(\th_1)|$ increases. This is illustrated in the right plot of  Fig.~\ref{fdec} for $S=S_{\th_1}S_{-\th_1^*}S_{\th_2}S_{-\th_2^*}$ with $\th_1=3+i\pi/7$ and $\text{Im}\, \th_2= i \pi/9$ held fixed.  The black line vanishes at a positive value of $S^\prime(\th_1)$ because below a critical coupling there is no bound on the mass gap for a system with only two resonances $\{\th_1, \, \th_2\}$. Further assumptions on the spectrum of possible  higher mass resonances would lead to stricter bounds on the separation $\text{Re}\th_1- \text{ Re} \th_2$.

\subsection{Interpretation of the bound}
\label{inter}

Consider an effective action describing the low energy dynamics of two massive scalar fields with a cubic interaction in two dimensions,
\begin{equation}
\label{paction}
	S= \int \rmd^2x\, \left[\frac{1}{2}\left(\partial^{\mu}\pi\partial_{\mu}\pi-m^{2}\pi^{2}+\partial^{\mu}\sigma\partial_{\mu}\sigma - M^{2}\sigma^{2}\right)-\frac{\lambda}{2}\sigma\pi^{2}-\dots\right]\,,
\end{equation}
where $\cdots$ denote further interactions of the fields that stabilize the potential at large field values but whose coupling constant is much smaller than $\lambda/m^{2}$ and are therefore inconsequential for the discussion below.

\subsubsection{Perturbative $S$-matrix}
Due to the cubic vertex in \eqref{paction}, for $M>2m$ the particle excitations of $\sigma$ are unstable and can decay to lighter particles. This instability manifests itself as a resonance in the $\pi\pi\rightarrow\pi\pi$ scattering which can be analyzed with perturbation theory. The $\pi\pi\rightarrow\pi\pi$ component of the $S$-matrix is given by\footnote{See appendix \ref{pertdetails} for details.}
\begin{equation}
	\mathbf{S} = S(s)\,\mathds{1} = \left(1+ \frac{i \cM(s)}{2\sqrt{s}\sqrt{s-4m^{2}}}\right) \mathds{1} \label{psmat}
\end{equation}
where the identity is the inner product of two particle states $\mathds{1} = \langle p_{3},p_{4}|p_{1},p_{2}\rangle$.
The amplitude $\cM$ is given by the sum of the $\sigma$-exchange diagrams in the $s,t$ and $u$-channel:
\be
	i\cM(s)=
		\begin{minipage}[h]{0.1\linewidth}
        \includegraphics[width=\linewidth]{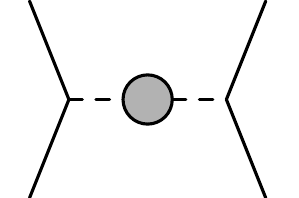} 
   \end{minipage}+
   	\begin{minipage}[h]{0.1\linewidth}
        \includegraphics[width=\linewidth]{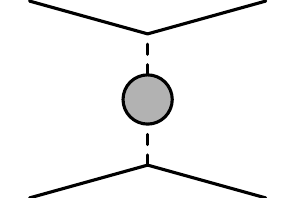}
   \end{minipage}+
   	\begin{minipage}[h]{0.1\linewidth}
        \includegraphics[width=\linewidth]{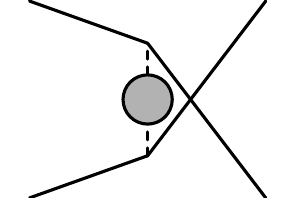}
   \end{minipage} \ .
\ee
Up to higher order loop corrections and non-perturbative effects $\cM$ is given by
\begin{equation}
\label{amp}
	{\cal M}(s) = \frac{\lambda^{2}}{M^{2}+ \frac{\lambda^{2}}{8\pi m^{2}}} - \frac{ \lambda^{2}}{s-M^{2}-\Pi(s)} - \frac{\lambda^{2}}{4m^{2}-s-M^{2}-\Pi(4m^{2}-s)} \, ,
\end{equation}
where $\Pi(s)$ is the (amputated) two-point function given by
\begin{equation}
	\Pi(s) = \frac{\lambda^{2}}{2\pi}\,\frac{\tanh^{-1}\sqrt{\frac{s}{s-4m^{2}}}}{\sqrt{s(s-4m^{2})}} \, . 
\end{equation}
As discussed in section~\ref{pwp}, we find that \eq{psmat} is crossing symmetric $S(s)=S(4m^2-s)$, it is  analytic in the domain $0<|s\pm i \eps |<4m^2$ and real along the real line in that domain.
At the two-particle production threshold $s=4m^2$ there is a square-root branch point and by crossing symmetry we find another one at $s=0$. 

It is convenient to resolve the square-root singularities at $s=0,4m^2$ by means of the conformal map in \eq{conf1}. Henceforth we work in units such that $m=1$. As a function of $\th$, the $S$-matrix in (\ref{psmat}) is single-valued and given by
\be
	S(\theta) = 1+\frac{i\lambda^{2}}{4\sinh\theta} \Bigg[\frac{1}{M^{2}+ \frac{\lambda^{2}}{8\pi}} - \frac{1}{4\cosh^{2}\left(\frac{\theta}{2}\right)-M^{2}-\Pi(\theta)} 
+ \frac{1}{4\sinh^{2}\left(\frac{\theta}{2}\right)+M^{2}+\Pi(i\pi-\theta)}\Bigg]
\label{psmatth} \, ,\ee
where the amputated two-point function is given by
\be
	\Pi(\theta)= \frac{\lambda^{2}}{8\pi} \frac{\theta-i\pi}{\sinh{\theta}} \, , \label{ptth}
\ee
for $\th$ in the fundamental domain $\th\in(-i\pi,i\pi]$.~\footnote{\eq{ptth}  is valid in the fundamental domain $\theta\in(-i\pi,i\pi]$. However, it can  be extended to the entire complex $\theta$-plane by means of the identity $\th-i\pi=\log\left(\frac{\sinh\th/2-\cosh\th/2}{\sinh\th/2+\cosh\th/2}\right)$, valid in the  fundamental domain,   where  the r.h.s. is periodic in the imaginary axis direction with period $\theta\sim\theta+2\pi i$ .} 

\subsubsection{Zeros and poles}
For $M>2$ the $S$-matrix in \eq{psmatth} has four poles and four zeros. The two $s$-channel poles are located at
\be
	\theta_{p}^{\pm} =\pm \left(\theta_{\circ}+\frac{\lambda^{2}\theta_{\circ}}{16\pi\sinh^{2}\theta_{\circ}}\right) - \frac{i\lambda^{2}}{16 \sinh^{2}\theta_{\circ}} \, ,
\ee
where $\theta_{\circ}=2\cosh^{-1}\left({\frac{M}{2}}\right)$, and the two zeros are located at
\be
	\theta_{z}^{\pm} = \pm \left(\theta_{\circ}+\frac{\lambda^{2}\theta_{\circ}}{16\pi\sinh^{2}\theta_{\circ}}\right) + \frac{i\lambda^{2}}{16 \sinh^{2}\theta_{\circ}}
	\label{pz} \, , 
\ee
in the physical strip.  Note that 
$	\theta^{\pm}_{z} = - \theta^{\mp}_{p}
$, as required by unitarity. 
The location of the remaining poles and zeros follow from crossing symmetry, \emph{i.e.} $S(\th)=S(i\pi-\th)$, and are located at 
\be
	\theta_{p/z}^{c\, \pm}= i\pi - \theta_{p/z}^{\pm} \, . 
\ee
Let us remark that the zeros  $\{\th_z^\pm, \th_{z}^{c\, \pm}\}$ lie above the real axis corresponding to the physical sheet while the poles $\{\th_p^\pm, \th_{p}^{c\, \pm}\}$ lie below the real axis corresponding to the second Riemann sheet of the $s=0,4m^2$ branch points.
Thus, the picture is qualitatively similar to the one depicted in Fig.~\ref{figex}.

\subsubsection{Bound}

In order to better understand the scope of the bound \eqref{lgp2} lets consider it in the context of the perturbative example. 
Near the root at $\theta=\theta_{i}$ one of the denominators in the perturbative $S$-matrix \eqref{psmatth} is of order $\lambda^{2}$, see appendix \ref{pertdetails} for details. Consequently the dominant contribution to $S^{\prime}(\theta_{i})$ goes like $\lambda^{-2}$. For $\theta_{i}=\theta^{\pm}_{z}$ given by \eqref{pz} we find
\begin{equation}
	S^{\prime}(\theta_{i}) = -\frac{2i}{\lambda^{2}} M^{2}(M^{2}-4) + O(\lambda^{0}) \, ,
\end{equation}
thus \eq{lgp2} bounds $\sim M^4/\lambda^2$ (in units of $m=1$).
The width of the resonance associated to $\theta_{i}$ is given by
\be
	\Gamma = -\Im \Pi(\theta_{\circ})/M =\lambda^{2}/(4M^{2}\sqrt{M^{2}-4})
	\ee
	 and hence   $S^{\prime}(\theta_{i})$ can be expressed as
\begin{equation}
	|S^{\prime}(\theta_{i})| = \frac{\lambda^{2}}{8M^{2}\Gamma^{2}} \, .  \label{xpert}
\end{equation}
Note that at this order in perturbation theory the effective perturbative parameter   is given by $\lambda_{\text{eff}}=\frac{\lambda}{M}$ (as can be seen in (\ref{psmatth})). In light of this observation, \eq{xpert} can be interpreted as follows: for a resonance particle $\sigma$ with a fixed width $\Gamma$, the strength of the coupling between $\sigma$ and the stable particles $\pi$ is constrained to obey bound \eqref{lgp2}. This indicates that for fixed width, $|S^{\prime}(\theta_{i})|$ is directly related to the coupling parameter of the stable particles to the resonance particle $\theta_{i}$ and can therefore be (loosely) referred to as the coupling. 
We remark that the bound is  saturated at this order of perturbation theory, and that  higher order corrections set $|S^\prime(\th_i)|$    within the bound due to particle production $S_{2\rightarrow 4}>0$.

\section{ Numerical optimization}
\label{nums}

The results presented in the previous section do not admit a straightforward generalization to higher dimensions where the analytical approach proves cumbersome. For this reason here we present an alternative approach utilizing numerical methods which can be generalized to higher dimensions. This approach can be summarized in two key steps. First, we construct an ansatz for the $S$-matrix which encodes analyticity and crossing symmetry as well as the location of the unstable resonances.   The free parameters of the ansatz include  the resonance coupling parameters. The second step is to maximize these coupling parameters under the constraint of unitarity $S(\th)S(-\th)\leq 1$ thus recovering the analytical bound of the previous section.

In $1+1$ dimensions the $2\rightarrow 2$ $S$-matrix element admits a simple expansion which we can exploit to build our ansatz.  Although this ansatz does not generalize to higher dimensions it serves as a simple framework to demonstrate how the numerical approach outlined above is implemented.  Thus we first present the numerical approach using this ansatz before presenting a more general ansatz which can readily be generalized to higher dimensions.

\begin{figure}[t]
\be
\begin{minipage}[h]{0.1\linewidth}
\includegraphics[scale=.475]{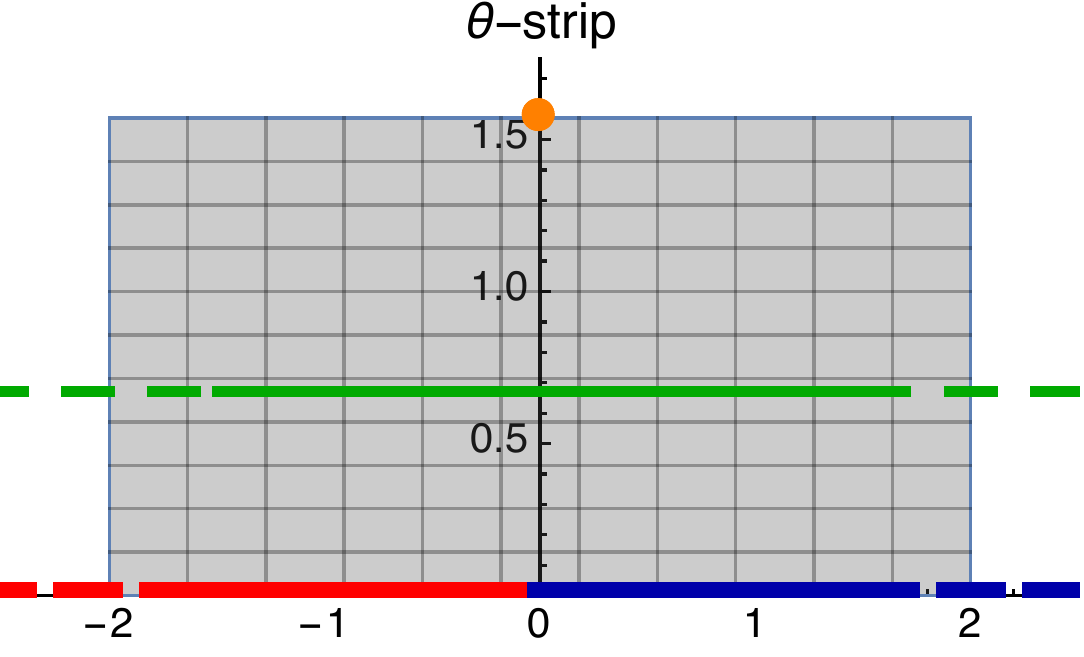}
\end{minipage} 
\hspace{4.cm}
\begin{minipage}[h]{0.1\linewidth}
\includegraphics[scale=.3475]{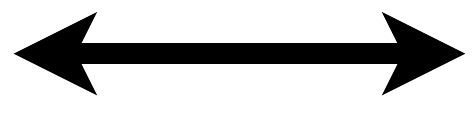}
\end{minipage}
\hspace{.5cm}
\begin{minipage}[h]{0.1\linewidth}
\includegraphics[scale=.475]{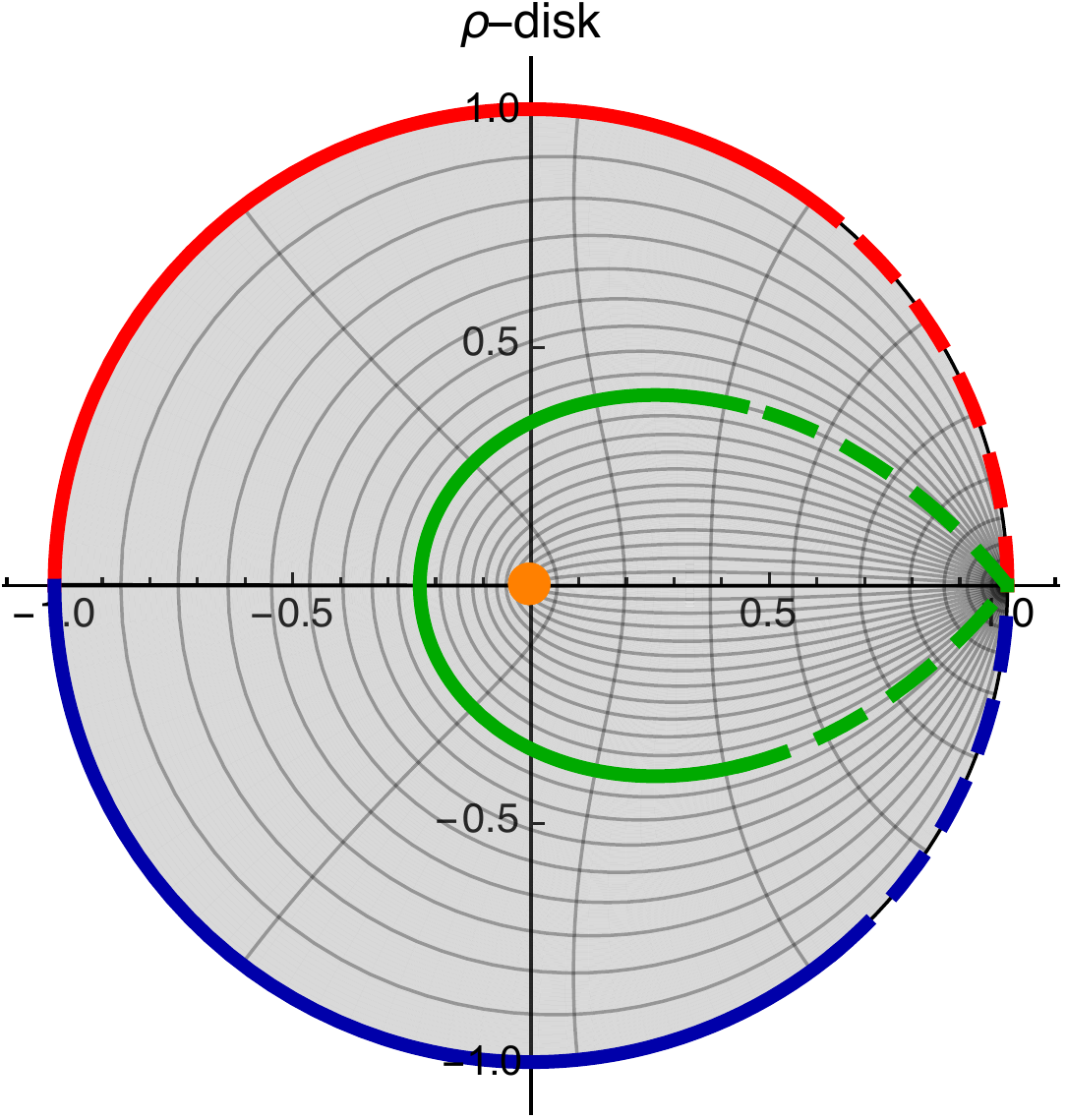} 
\end{minipage} \hspace{3cm} \nonumber
\ee
\caption{Illustration of the conformal map in \eq{map} with $\beta=i\pi$. \label{conf2}} 
\end{figure}


\subsection{A simple $S$-matrix ansatz for $d=1+1$}
\label{1ans}

Let us denote by $S_\text{ans}^{\{\th_j\}}$ our ansatz for the $2\rightarrow 2$ $S$-matrix element. This ansatz is labelled by the location of its roots in the $\theta$-strip $\{\th_j\}$. To encode holomorphicity of  the function $S_\text{ans}^{\{\th_j\}}$ in a domain $\cD_\th$ of the $\th$-strip we consider a conformal map $\rho$ from $\cD_\th$ into the unit disk. The $S$-matrix, viewed as a function of $\rho$, is therefore holomorphic inside the unit disk. A  holomorphic function inside the unit disk is analytic and therefore admits an absolutely convergent Taylor expansion inside the disk. Thus $S_\text{ans}^{\{\th_j\}}(\rho)$ can be defined via its Taylor expansion inside the unit disk which makes holomorphicity of $S_\text{ans}^{\{\th_j\}}(\rho(\th))$ in $\cD_\th$ manifest.

To make crossing symmetry $S_\text{ans}^{\{\th_j\}}(\th)= S_\text{ans}^{\{\th_j\}}(i \pi-\th)$ manifest we require the map $\rho(\th)$ to satisfy $\rho(\th)=\rho(i\pi-\th)$. Such a map can be viewed as a biholomorphic map between the fundamental domain $\Im\th\in(0,\pi/2)$ and  the unit disk. A conformal map with such properties is given by  
\be
\rho(\th)=\frac{\sinh\th-i}{\sinh\th+i} \, . \label{map}
\ee
As illustrated in Fig.~\ref{conf2}, under the map \eq{map} the crossing symmetric point $\th=i\pi/2$ is mapped into the origin of the disk while the real axis is mapped to the boundary.

Since we require that the ansatz   vanishes at $\rho_{\th_i}=\rho(\th_i)$,   the function 
\be
g(\rho)\equiv S^{\{\th_j \}}_\text{ans}(\rho) / \prod_j (\rho-\rho_{\th_j})  \label{comb1}
\ee
is nowhere vanishing and holomorphic inside the unit disk and therefore admits an absolutely convergent Taylor expansion
\be
\label{comb2}
g(\rho)=  z  \, (1+ \sum_{n=1}^{\infty}c_n \rho^n) \,,
\ee 
where the overall factor is given by $z\equiv g(0)>0$.

Combining (\ref{comb1}) and (\ref{comb2}) results in the following representation of our ansatz $S$-matrix element
\be
S^{\{\th_j \}}_\text{ans}(\rho)=   \, z  \,  \prod_j (\rho-\rho_{\th_j})  \, \big(1+  \sum_{n=1}^\infty c_n \rho^n \big) \, , \label{ans}
\ee
which is holomorphic, crossing symmetric and encodes the location of the zeros $\rho_{\theta_{j}}$. The parameters  $\{ z, c_i \}$ are constrained by unitary of the $S$-matrix \eq{unitrho} but are otherwise  arbitrary real parameters.

Our next step is to maximize $z$ over the space of the expansion coefficients $\{c_{n}\}$ in \eq{ans} under the constraint of unitarity. The unitarity bound along the real line in the $\th$-strip translates to an analogous bound  along the boundary of the unit disk parameterized as $\rho=e^{i\phi}$,
\be
S^{\{\th_j \}}_\text{ans}(e^{i\phi})\, S^{\{\th_j \}}_\text{ans}(e^{-i\phi})�\leq 1   \quad \text{for}\quad \phi\in[0,\pi] \label{unitrho} \, . 
\ee
Maximizing over $z$   is tantamount to maximizing over the coupling to the resonances $\partial_{\th_k}S_\text{ans}^{\{\th_j\}}(\rho_{\th_k})$, with the location of the roots of $S_\text{ans}^{\{\th_j\}}$ held fixed. All couplings  are maximized simultaneously, as follows from (\ref{lgp2}).

\subsubsection{Numerical  results}

In order to set up the numerical code to maximize over $z$, the series in (\ref{ans}) has to be truncated. Therefore in the numerical code we maximize $z$ in the truncated ansatz
\be
S^{\{\th_j \}}_\text{M}(\rho)=   \, z  \,  \prod_j (\rho-\rho_{\th_j})  \, \big(1+  \sum_{n=1}^M c_n \rho^n \big)\, .  \label{nans}
\ee
In addition, the constraint (\ref{unitrho}) is imposed in  a large but finite number of points on the unit circle, namely it is   evaluated at  $K$  points 
\be
\phi \in \left\{0,\pi/K,2\pi/K,\dots , \pi  \right\} \, . \label{pts}
\ee 
The only approximation made in this implementation is in the truncation of the series in (\ref{nans}). Convergence as $M$ is increased is fast and even keeping the first few terms leads to precise results.

\begin{figure}[t]\centering
\includegraphics[scale=.675]{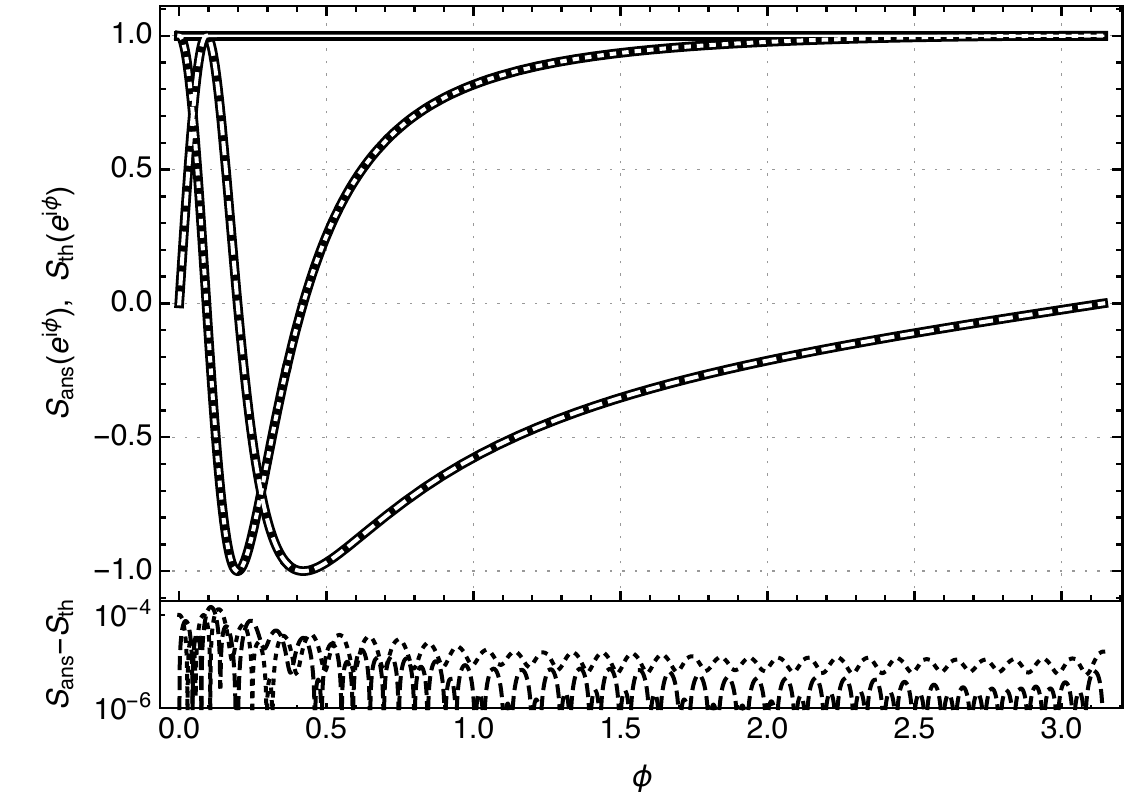} \hspace{.5cm}
\includegraphics[scale=.675]{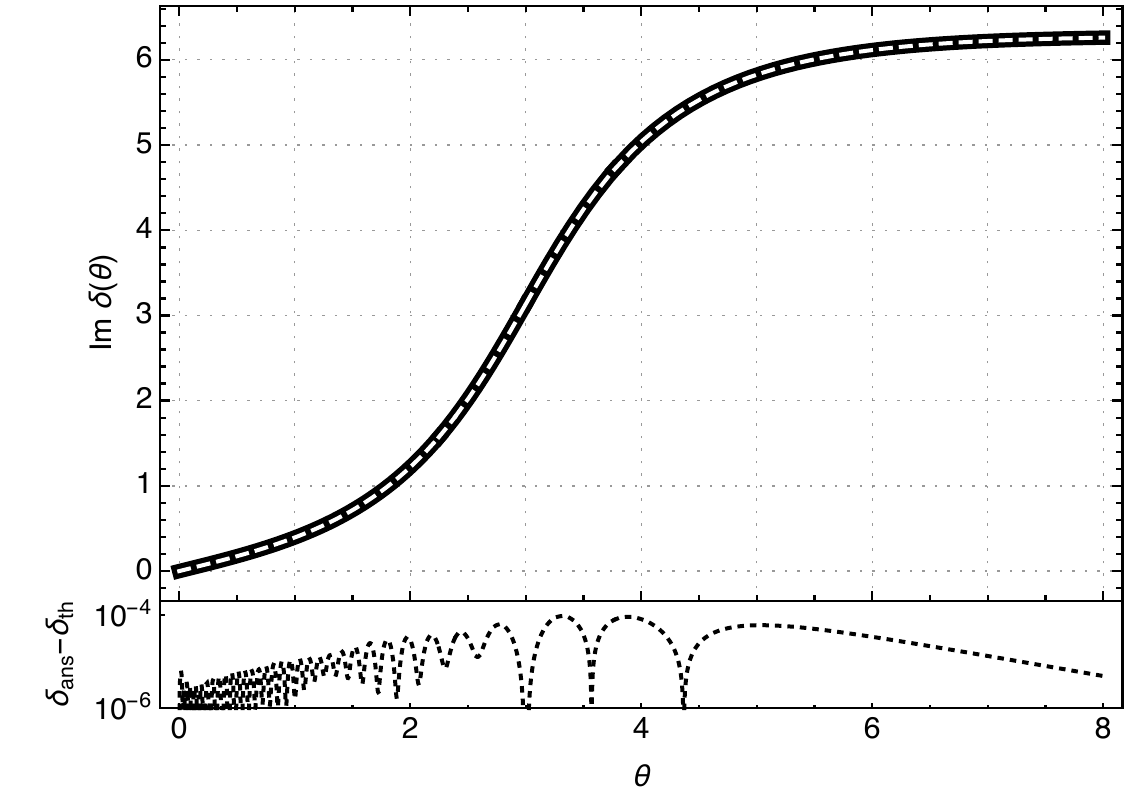}
\caption{In the left plot solid black lines depict the imaginary part, real part and absolute value of $S_\text{th}(e^{i\phi})$. Superimposed we show the imaginary part (dashed white), the real part (doted white) and absolute value (solid white) of $S_\text{ans}(e^{i\phi})$. In the lower left plot, sharing the same horizontal axis, we show the real part (dotted) and the imaginary part (dashed) of $S_\text{th}(e^{i\phi})-S_\text{ans}(e^{i\phi})$. Finally the plot to the right is a comparison of the phase of   $S_\text{ans}$ (white) and $S_\text{th}$ (black) in the $\th$ strip.  \label{num1}} 
\end{figure}

As an example, consider 
\be
\{\th_j\} = \{3+i, -3+i\}  \, ,\label{szeros}
\ee
in \eq{nans}. We maximize (\ref{nans}) over $z$ under the unitary constraint   for the set of zeros in (\ref{szeros})  and we get the white lines depicted in Fig.~\ref{num1}.  To obtain  such result we increased   $M$ and $K$ until we got a convergent result, the plots shown are for $\{M,K\}=\{60,90\}$. 
The white lines are super-imposed over black thicker lines. These correspond to the $S$-matrix theory $S_\text{ex}$ of (\ref{scatex}) with $\alpha\rightarrow 3+i$
\be
S_\text{ex}(\th)=S_{3+i}(\th) S_{-3+i} (\th)  \, ,  \label{ths}
\ee
in the left plot, while in the right plot we compare with the phase-shift of $S_\text{ex}$.\footnote{
Recall that $S_\text{ex}$ saturates unitarity at all physical energies. In $d=3+1$, only the trivial $S$-matrix saturates unitarity at all energies, since any non-trivial $S$-matrix has a finite amount of particle production.  From this perspective, wether or not we can identify a   Lagrangian model leading to  $S_\text{ex}$ is somewhat anecdotic and special to $d=1+1$ physics. } 
  The numerical results have been  obtained with \texttt{Mathematica}'s function  \texttt{FindMaximum}. The computation is cheap, taking $O(1)$~min. of time and $\sim 1$~Mb  of memory RAM. 
Optimization of  $S$-matrices with many more resonances is also feasible and leads to equally good results.

\subsection{Towards  generalization to higher dimensions}

It is convenient to make crossing symmetry explicit by extending $S$ into a symmetric function of two Mandelstam variables $S(s,t)=S(t,s)$. An ansatz of this form is much more suitable for generalization to higher dimensions. In $1+1$ dimensions the physical $S$-matrix is obtained by constraining $S(s,t)$ to the plane $s+t=4m^2$. This function is analytical in $s$ and $t$ up to the branch points on the real line. 

Next,  we   build an ansatz  $S_{\text{ans}-2}^{\{\om_i\}}(s,t)$ encoding analyticity, crossing, and the location of the resonances. Similar to what we did in section~\ref{1ans} we encode analyticity in each variable $s$ and $t$  by conformally mapping the domain of holomorphicity into the unit disk and subsequently define the function as a Taylor series in the poly-disk. Such a conformal map is provided by
\be
 \om(s)= \frac{\sqrt{2m^2}-\sqrt{4m^2-s}}{\sqrt{2m^2}+\sqrt{4m^2-s}} \, .  \label{defw}
\ee
As we did before, we factor out the zeros of $S_{\text{ans}-2}^{\{\om_i\}}(s,t)$, and expand the nowhere vanishing part in a (convergent) double-expansion whilst fulfilling all the physical assumptions,
\be
S_{\text{ans}-2}^{\{\om_i\}}(s,t)=  z \prod_i \left[ \om_i-\om(s)\right]\left[ \om_i-\om(t)\right] \Big(  1+   \sum_{m,n=1}^\infty c_{n,m}\,  \om(s)^{n}\om(t)^{m} \Big)  \, .\label{toproof}
\ee
Here $c_{m,n}$ are symmetric and real. \eq{toproof} can be equivalently written in the corresponding $\th_s,\, \th_t$-strips.
The existence of such a double expansion \eq{toproof} is easy to show in two dimensions. To this end note that the map \eqref{map}, with $\theta=2\cosh^{-1}(\frac{\sqrt{s}}{2m})$, has the following convergent expansion
\begin{equation*}
	\rho(s)=  \left[\om(s)+\om(4m^{2}-s)\right]  \sum_{n=0}^\infty (-1)^n  \, \om(4m^2-s)^{n}\, \om(s)^{n}\,.
\end{equation*}
This, together with \eqref{ans}, results in the double expansion \eqref{toproof} with $t=4m^{2}-s$.


Evaluating  $S_{\text{ans}-2}^{\{\om_i\}}(s,t)$ at $t=4m^2-s$, and maximizing over $z$ under the unitarity constraint we obtain comparable results to the ones showed in Fig.~\ref{num1}. 
\eq{toproof} admits a  generalization from $d=1+1$ to $d=3+1$ spacetime dimensions as was done in \cite{Paulos:2017fhb}. In $d=3+1$ there is an analogous definition of resonance presented in section  \ref{class}, in terms of phase-shifts and roots of the components of the $S$-matrix in the partial wave decomposition. In a forthcoming publication we plan to study    the space of $S$-matrices in $d=3+1$ that feature unstable resonances.

%

\section{Summary and outlook}

In this work we have found a bound on the coupling of asymptotic states to unstable resonances,  \eq{lgp2}, which is saturated in the limit of maximal elasticity of the $2\rightarrow 2$ $S$-matrix element.
The bound for each resonance is improved as the number of resonances is increased, or the gap between the resonances is decreased. Therefore (\ref{lgp2}) can be interpreted as setting a minimal mass gap between the resonances as a function of the coupling to the resonances. 
In section~\ref{nums} we have recovered the analytical results of section~\ref{main1} as a numerical solution. This consists in numerically maximizing the coupling to the resonances of the $S$-matrix ansatz \eqref{nans} or \eqref{toproof} under the constraint of unitarity.

There are a number of interesting directions left to be developed. For instance, in $d=1+1$ spacetime dimensions, generalizing our results to systems involving many non-trivial $2\rightarrow 2$ $S$-matrix elements is of interest and could prove instructive for more involved problems in higher dimensions. This would also facilitate making contact with integrable models such as ref.~\cite{Miramontes:1999hx} which have a known Lagrangian description and feature unstable particles. Another interesting direction in $d=1+1$ is to constrain unstable resonances by studying the crossing symmetry constraints on four-point functions in the boundary of  AdS in $d=1+1$ and subsequently taking the flat space limit~\cite{Paulos:2016fap}.


Perhaps the most promising direction to pursue, from the particle physics point of view, is to generalize the results obtained here to $d=3+1$. In section~\ref{nums} we have explained a possible route towards such generalization. This avenue promises many applications to particle physics and model building beyond the Standard Model. It would also be interesting to study unstable particles of higher spin in the $2\rightarrow 2$ scalar $S$-matrix element which can be achieved via incorporating the corresponding Legendre polynomials in the $S$-matrix ansatz.
 
 A naive generalization of the bound \eqref{lgp2} to $d=3+1$ suggests that we should find a maximal value of $|S'|$ for  unstable resonances which is saturated in the limit where the amount of particle production is minimized. Furthermore, we anticipate an interesting interplay between the maximal value of $|S'|$  and the number of resonances allowed below a given energy. For instance, in analogy to the result of section~\ref{morecom}, in $d=3+1$ we expect that given the value  $S'(\th_i)$ of the lightest resonance there is a lower bound on the mass of the next-to-lowest lying resonance.  

\subsection*{Acknowledgements}

We thank G.~Mussardo,  
 S.~Rychkov,
  B.~van Rees and 
 G.~Villadoro for the useful discussions.  We are also grateful to   J.~Penedones, M.~Serone and L.~Vitale for the useful discussions and comments on the draft. N.~D. is supported by the PRIN project ``Non-perturbative Aspects Of Gauge Theories And Strings''.

\appendix

\begin{figure}[t]\centering
\includegraphics[scale=.17]{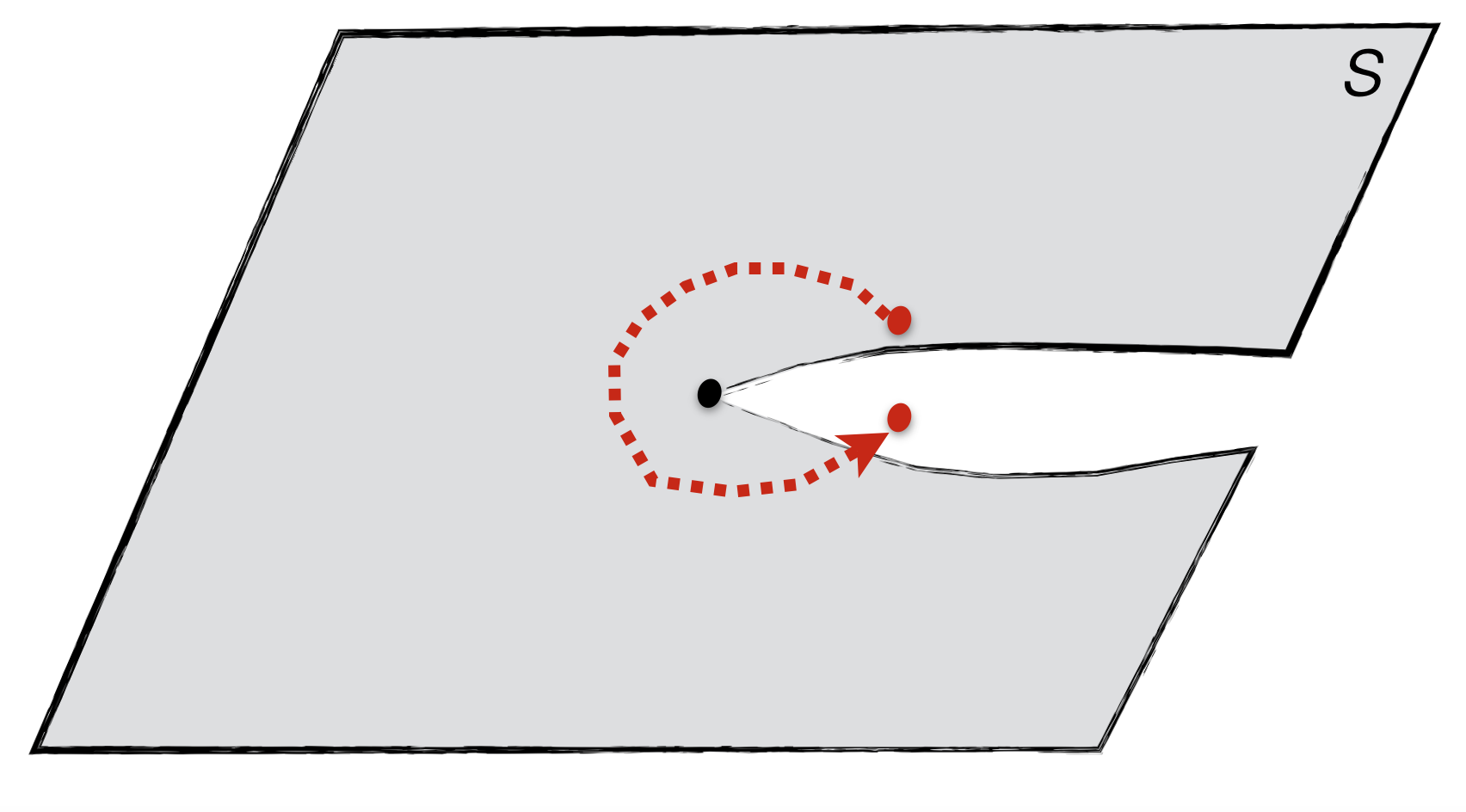}
\caption{Analytic continuation of  \eq{unit2} through the branch cut.  \label{conte} }
\end{figure}

\section{Nature of the two-particle branch point}
\label{branchpoint}


 \eq{unit2} can be  used to argue that the two-particle threshold branch point is a square-root  singularity, see for instance \cite{Mussardo:2010mgq}.
  The argument goes as follows. Consider the analytical continuation of $S(s+i\eps)$ into the second Riemann sheet by following a full anti-clockwise rotation around $s=4m^2$, see Fig.~\ref{conte}. Lets call such analytically continued function $G(s)$. Then, under such analytical continuation, the unitary equation becomes 
   \be
  S(s-i\eps) G(s-i\eps)=f(s) \, , \label{unit4}
 \ee
 where we made use of continuity $S(s-i\eps)=G(s+i\eps)$ (and assumed that $f(s)$ has no branch points).
Then, by taking the ratio between \eq{unit2} and \eq{unit4} one obtains
\be
G(s-i\eps)  =S(s+i\eps) \, .
\ee
The latter equation  would imply that rotating around the two-particle branch point twice the $S$-matrix is invariant. Therefore,   if the $S$-matrix has a branch point and if the branch point is an algebraic singularity then it must be a square-root type singularity.\footnote{Note that  this  argument alone can not exclude the possibility of two-sheeted essential singularities.}
A similar result can be obtained in $d=3+1$ dimensions~\cite{Eden:1966dnq}.

 \newpage

\section{Total integrated phase-shift}
\label{deri}

The function $S(\th)$ is meromorphic inside the physical strip $\Im\th\in(0,\pi)$. The so called \emph{argument principle} implies~\footnote{See for instance Theorem 4.1 of \cite{steinShakarchi}.} 
\be
\oint_{\partial P}\frac{d\th}{2\pi i}\, \partial_\th(2i \delta(\th))  = N_z-N_p \, . \label{numnp}
\ee
where the closed contour integral encircles some region $P$ in the physical strip and $N_z$ and $N_p$ are the number of zeros and poles in that region, weighted by their order. In our particular physical set up $P$ is inside the physical $\th$-strip and thus there are no poles ($N_p=0$) due to stable particles.

Now, using crossing symmetry we have $\partial_\th\delta(i\pi+\th)=-\partial_{\th}\delta(-\th)$ and therefore the   total phase-shift can be written as a contour integral
\be
 \int_{-\infty}^\infty d\th \, \partial_\th \delta(\th)   =  \frac{1}{2}\oint_{\cC} d\th \, \partial_\th \delta(\th) \, ,  \label{cont1}
\ee 
with $\cC$ is a contour encircling the whole physical strip. Then, by Cauchy residue theorem, \eqref{totalps} follow from \eqref{cont1}. For simplicity we have assumed that $S(\th)$ asymptotes to a constant at $\th\rightarrow\pm\infty$ so that  the contribution from the segments at infinity vanishes -- this assumption can be relaxed.

\section{Perturbative example}
\label{pertdetails}

In this appendix we provide further details of the perturbative QFT discussed in section \ref{inter}. 
The Feynman rules for the  theory in (\ref{paction}) are
\be
\begin{minipage}[h]{0.08\linewidth}
        \includegraphics[width=\linewidth]{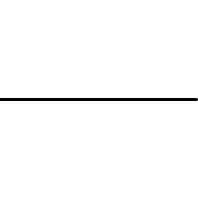}
   \end{minipage}=  \frac{i}{k^{2} - m^{2}+i\epsilon}
	 \ \ , 	\quad
	\begin{minipage}[h]{0.08\linewidth}
        \includegraphics[width=\linewidth]{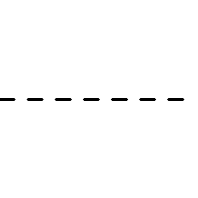}
   \end{minipage} 	=  \frac{i}{k^{2} - M^{2}+i\epsilon}
\	\ , 	\quad
	\begin{minipage}[h]{0.08\linewidth}
        \includegraphics[width=\linewidth]{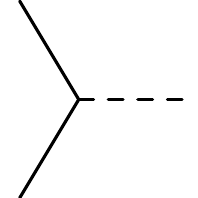}
   \end{minipage} =   -i\lambda \ \ , 
\ee
where the plain line denotes the propagator for $\pi$ and the dashed line denotes the propagator for $\sigma$. First order of business is to compute the loop corrections to propagation of $\sigma$ which is captured by the (amputated) diagram
\be
	-i\Pi(k^{2}) \, = \, 
	\begin{minipage}[h]{0.07\linewidth}
        \includegraphics[width=\linewidth]{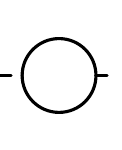}
   \end{minipage}
\ee
The loop integral can be carried out explicitly and yields 
\be
\label{pido}
	\Pi(k^{2}) = \frac{i}{2} (-i\lambda)^{2} \int\frac{\rmd^{2} q}{(2\pi)^{2}}\, \frac{i}{q^{2}-m^{2}+i\epsilon}\,\frac{i}{(k-q)^{2}-m^{2}+i\epsilon}= \frac{\lambda^{2}}{2\pi}\,\frac{\tanh^{-1}\sqrt{\frac{k^{2}}{k^{2}-4m^{2}}}}{\sqrt{k^{2}}\sqrt{k^{2}-4m^{2}}} \, ,
\ee
where the limit $\eps\rightarrow 0$ was taken in the integrated function.
Along the real axis and below the two-particle threshold $k^{2}< 4m^{2}$ the loop correction $\Pi(k^{2})$ is real. Above the two-particle threshold  $\Pi(k^2)$   has a non-vanishing imaginary part
\begin{equation}
	\Im\, \Pi(k^{2}) = -\frac{\lambda^{2}}{4\sqrt{k^{2}(k^{2}-4m^{2})}}\, \theta(k^{2}-4m^{2}) \, .  \label{impart}
\end{equation}
The latter equation can be extracted directly using Cutkosky rules.

\subsection{The perturbative $S$-matrix}

Incorporating the loop correction to the propagator \eqref{pido}, the quantum corrected propagator for $\sigma$ is given by
\be
	\cG(k^{2}) \, = \, 
	\begin{minipage}[h]{0.1\linewidth}
        \includegraphics[width=\linewidth]{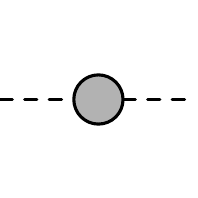}
   \end{minipage}
   =\frac{i}{k^{2}-M^{2}-\Pi(k^{2})+i\epsilon}
\ee
With this propagator we can readily compute the $\pi\pi \rightarrow \pi\pi$ component of the $S$-matrix. The contributing diagram in the $s$-channel is
\be
	\begin{minipage}[h]{0.13\linewidth}
        \includegraphics[width=\linewidth]{figs/s-channel.eps}
   \end{minipage} = \frac{-i \lambda^{2}}{s-M^{2}-\Pi(s)+i\epsilon} \ =\ -\lambda^{2}\cG(s)
\ee
where $s=(p_{1}+p_{2})^{2}$. In the centre of mass frame we have $p_{1}=p_{4}=(p^{0},p^{1})$ and $p_{2}=p_{3}=(p^{0},-p^{1})$. Thus $t=(p_{1}-p_{3})^{2}=4m^{2}-s$ and $u=(p_{1}-p_{4})^{2}=0$. The corresponding amplitude in the $t$-channel amounts to
 $
	-\lambda^{2}\cG(t)=-\lambda^{2}\cG(4m^{2}-s)
$
and the $u$-channel diagram yields a contribution equal to $-\lambda^{2}\cG(u)=-\lambda^{2}\cG(0)$. The one-loop amplitude -- up to higher order corrections~\footnote{Namely, vertex corrections and box diagrams.} -- is given by
\begin{equation}
	i\cM(s) = \frac{i\lambda^{2}}{M^{2}+ \frac{\lambda^{2}}{8\pi m^{2}}} - \frac{i \lambda^{2}}{s-M^{2}-\Pi(s)} - \frac{i \lambda^{2}}{4m^{2}-s-M^{2}-\Pi(4m^{2}-s)} \, .
\end{equation}
Then, the $\pi\pi\rightarrow\pi\pi$ component of the $S$-matrix is given by
\begin{equation}
\label{psmatrix}
	\mathbf{S} = S(s)\,\mathds{1} = \left(1+ \frac{i\cM(s)}{2\sqrt{s(s-4m^{2})}}\right) \mathds{1}
\end{equation}
where the identity is the inner product of two particle states $\mathds{1} = \langle p_{3},p_{4}|p_{1},p_{2}\rangle$, which is given by
\begin{equation}
	\mathds{1} = (2\pi)^{2} 4 E_{1}E_{2}\left(\delta(p_{1}^{1}-p_{3}^{1})\delta(p_{2}^{1}-p_{4}^{1})+\delta(p_{1}^{1}-p_{4}^{1})\delta(p_{2}^{1}-p_{3}^{1})\right) \, , 
\end{equation}
and the extra factor multiplying $\cM(s)$ arises from the  identity
$
	(2\pi)^{2} \delta^{(2)}(p_{1}+p_{2}-p_{3}-p_{4}) = \mathds{1}/(2\sqrt{s}\sqrt{s-4m^{2}})
$.

A few remarks regarding $S(s)$ defined in \eqref{psmatrix}
are in order. As we will demonstrate below, the above $S$-matrix has complex poles hidden behind a branch cut stemming from the two-particle threshold. These poles correspond to the exchange of an on-shell unstable $\sigma$ particle which amounts to a resonance in the scattering process whose width is determined by
\begin{equation}
	\Gamma = -\Im\Pi(M_{\rm ren}^{2})/M_{ren}
\end{equation}
The resonance is observed as a phase shift due to a branch cut in $\log S(s)$ stemming from such poles. Alternatively we can extract the same information about the resonance from the zeros of $S(s)$ as they too give rise to a branch cut in $\log S(s)$ and thus result in a phase shift. Moreover the zeros are in a sense more fundamental as they are not hidden behind any branch cuts and we do not need to analytically continue the $S$-matrix through branch cuts to study them. Below we analyze the resonance in our perturbative example first by studying the poles and later through the zeros.

\subsection{Resonances, poles and zeros}
\label{zp}

As discussed in section \ref{class}, resonances are associated with branch cuts in the $s$-plane for the function $\log S(s)$ which typically connect a pair of pole and zero of the $S$-matrix the location and residue of which determine the features of the resonance. Here we study the zeros and poles of the perturbative $S$-matrix \eqref{psmatrix}.
\subsubsection{Poles in $s$-plane}
Poles associated to unstable resonances are not visible on the complex $s$-plane and are hidden behind multi-particle branch cuts of the $S$-matrix. To illustrate this in our perturbative example note that \eqref{psmatrix} has square-root branch points at $s=0$ and at $s=4m^{2}$. We take the branch cuts to stretch along the real axis from $-\infty$ to $0$ and from $4m^{2}$ to $+\infty$.  The location of the poles of $S(s)$ due to the exchange of a $\sigma$ in the $s$-channel are determined by
\be
\label{spoles}
	s-M^2- \Pi(s)=0 \, .
\ee
Now consider the following ansatz for the solution 
\begin{equation}
\label{poleansatzs}
	s_{*}=4m^2 +s_{\circ}e^{i\varphi+ 2\pi i n}
\end{equation}
where $s_{\circ}$ is taken to be positive, $\varphi\in[0,2\pi)$ and $n=0,1$ labels the sheets of the Riemann surface associated with the function $\sqrt{s-4m^{2}}$. Plugging our ansatz into the equation \eqref{spoles} and taking the real part we obtain
\begin{equation}
	s_{\circ} \cos \varphi + 4m^2-M^2-\text{Re}\Pi(s_{*}) =0 \, . 
\end{equation}
This implies $s_{*}=M^{2}+O(\lambda^{2})$ and since $4m^2-M^2-\text{Re}\Pi(M^2)<0$ we must have $\varphi\in(0,\pi/2)\cup(3\pi/2,2\pi)$. Therefore, the imaginary part of \eqref{spoles}   simplifies to
\begin{equation}
	s_{\circ} \sin \varphi +\frac{\lambda^{2}}{4} \frac{(-1)^{n}\cos(\varphi/2)}{M\sqrt{s_{\circ}}} =0  \, . 
\end{equation}
where we have used \eqref{impart} and have omitted higher order terms using that $\sin\varphi=O(\lambda^{2})$. It is evident that the equation has no solution for $n=0$, \emph{i.e.} in the physical sheet. To find a solution we have to take $n=1$ which  as can   be seen from \eqref{poleansatzs}  corresponds to the analytic continuation of the function $\sqrt{s-4m^{2}}$ into    the second sheet.~\footnote{Here we have assumed $M\ge 2m$. For $M<2m$ we find a pole on the real axis and below the two-particle threshold corresponding to production of a stable particle of mass $M$.}

Having demonstrated that the poles lie behind the square-root branch cut stemming from the two-particle threshold at $s=4m^{2}$ we now switch to the $\th$-variable, introduced in section \ref{introth}. Recall that  $\theta=\theta(s)$ maps the two-sheeted Riemann surface associated to \eq{psmatrix} into the strip $\theta\in[-i\pi,i\pi)$.

\subsubsection{Poles and zeros of the $S$-matrix}
 
 The $S$-matrix on the $\th$-strip, in units $m=1$, was given in (\ref{psmatth}): 
\be
	S(\theta) = 1+\frac{i\lambda^{2}}{4\sinh\theta} \Bigg[\frac{1}{M^{2}+ \frac{\lambda^{2}}{8\pi}} - \frac{1}{4\cosh^{2}\left(\frac{\theta}{2}\right)-M^{2}-\Pi(\theta)} 
+ \frac{1}{4\sinh^{2}\left(\frac{\theta}{2}\right)+M^{2}+\Pi(i\pi-\theta)}\Bigg] \, . 
\ee
We remind the reader that the above expressions are perturbative results valid only up to order $\lambda^{3}$. We are interested in finding poles and zeros of the $S$-matrix in the fundamental domain of complex $\theta$. To this end we can use a series expansion for small $\lambda$. The poles arise when one of the denominators vanishes. Since $S=1+O(\lambda^{2})$, the zeros lie in the regions where one of the denominators is of order $\lambda^{2}$. Thus we can look for location of zeros and poles in parallel. The location of the zeros and poles arising from the $s$-channel contribution are determined by the equation
\begin{equation}
\label{pandz}
	4\sinh{\theta}\,\left(4\cosh^{2}\left(\th/2\right)-M^{2}-\Pi(\th) \right)=ia\lambda^{2}
\end{equation}
where $a=0$ for poles and $a=1$ for zeros.

Note that the $t$ and $u$-channel contributions to \eqref{pandz} appear at order $\lambda^{4}$ along with contributions from other diagrams we have not considered. These contributions only affect the position of the pole and the zero at order $\lambda^{4}$. We are interested in poles and zeros near $s_{\circ}=M^{2}$ or $\theta_{\circ}^{\pm}=\pm2\cosh^{-1}\left({\frac{M}{2}}\right)$. We therefore look for perturbative solutions of the form
\begin{equation}
	\theta_{\sigma}^{\pm}= \theta_{\circ}^{\pm} + \lambda^{2} \theta_{*}^{\pm}
\end{equation}
Plugging this ansatz into \eqref{pandz} we find
\begin{equation}
	8\,\theta_{*}^{\pm}\,\sinh^{2}\theta_{\circ}^{\pm}-\frac{\theta_{\circ}^{\pm}-i\pi}{2\pi} = ia
\end{equation}
and therefore
\begin{equation}
	\theta_{*}^{\pm} =\frac{1}{8\sinh^{2}\theta_{\circ}^{\pm}} \left( \frac{\theta_{\circ}^{\pm}}{2\pi}+i(a-\frac{1}{2})\right)
\end{equation}
Thus we find two poles at
\begin{equation}
\label{thetap}
	\theta_{p}^{\pm} =\pm \left(\theta_{\circ}+\frac{\lambda^{2}\theta_{\circ}}{16\pi\sinh^{2}\theta_{\circ}}\right) - \frac{i\lambda^{2}}{16 \sinh^{2}\theta_{\circ}}
\end{equation}
and two zeros located at
\begin{equation}
\label{thetaz}
	\theta_{z}^{\pm} = \pm \left(\theta_{\circ}+\frac{\lambda^{2}\theta_{\circ}}{16\pi\sinh^{2}\theta_{\circ}}\right) + \frac{i\lambda^{2}}{16 \sinh^{2}\theta_{\circ}}
\end{equation}
where $\theta_{\circ}=2\cosh^{-1}\left({\frac{M}{2}}\right)$. Note that the zeros lie above the real axis corresponding to the physical sheet while the poles lie below the real axis corresponding to the second sheet and that the location of the zeros and poles are related by
\begin{equation}
	\theta^{\pm}_{z} = - \theta^{\mp}_{p}
\end{equation}
as required by unitarity. We also find a second pair of poles and zeros located at
\begin{equation}
	\theta_{p/z}^{c\, \pm}= i\pi - \theta_{p/z}^{\pm}
\end{equation}
using crossing symmetry.

We remind the reader that \eqref{thetap} and \eqref{thetaz} are only valid up to corrections of order $\lambda^{4}$ while \eqref{thetaz} only satisfy $S(\theta)=0$ up to corrections of order $\lambda^{2}$. This is due to the fact that near the roots of $S(\theta)$ the $s$-channel denominator is of order $\lambda^{2}$ resulting in higher order diagrams to contribute at order $\lambda^{2}$. On general grounds we expect the contribution of higher order diagrams to merely shift the location of the poles and zeros without affecting their order such that the $S$-matrix takes the form
\begin{equation}
	S(\theta) = \left(g(\theta) + O(\lambda^{2}) \right) \prod_{i} \frac{\th-\th_{z_i}}{\th-\th_{p_i}} \, , 
\end{equation}
where $\th_{z/p}$ are determined up to $O(\lambda^4)$.


\small

\bibliography{smb_jhep.bbl}
\bibliographystyle{utphys}

\end{document}